\RequirePackage{rotating}

\documentclass[twocolumn]{aastex61}
\usepackage{rotating}
\usepackage{capt-of}
\usepackage{amsmath}
\usepackage{tablefootnote}
\usepackage{natbib}

\shorttitle{Gaussian Process Star Formation Histories I: Major Episodes}
\shortauthors{Iyer et al.}

\begin{document}

\title{Non-parametric Star Formation History Reconstruction with Gaussian Processes I:  Counting Major Episodes of Star Formation}

\correspondingauthor{Kartheik Iyer}
\email{kgi1@physics.rutgers.edu}

\author[0000-0001-9298-3523]{Kartheik G. Iyer}
\affiliation{Department of Physics and Astronomy, Rutgers, The State University of New Jersey, 136 Frelinghuysen Road, Piscataway, NJ 08854-8019 USA}

\author[0000-0003-1530-8713]{Eric Gawiser}
\affiliation{Department of Physics and Astronomy, Rutgers, The State University of New Jersey, 136 Frelinghuysen Road, Piscataway, NJ 08854-8019 USA}
\affiliation{Center for Computational Astrophysics, Flatiron Institute, 162 5th Ave, New York, NY 10010, USA}

\author[0000-0003-4996-214X]{Sandra M. Faber}
\affiliation{University of California Observatories/Lick Observatory, University of California, Santa Cruz, CA 95064, USA}

\author[0000-0001-7113-2738]{Henry C. Ferguson}
\affiliation{Space Telescope Science Institute, 3700 San Martin Drive, Baltimore, MD 21218, USA}

\author[0000-0002-6610-2048]{Anton M. Koekemoer}
\affiliation{Space Telescope Science Institute, 3700 San Martin Drive, Baltimore, MD 21218, USA}

\author[0000-0003-4196-0617]{Camilla Pacifici}
\affiliation{Space Telescope Science Institute, 3700 San Martin Drive, Baltimore, MD 21218, USA}

\author{Rachel Somerville}
\affiliation{Center for Computational Astrophysics, Flatiron Institute, 162 5th Ave, New York, NY 10010, USA}
\affiliation{Department of Physics and Astronomy, Rutgers, The State University of New Jersey, 136 Frelinghuysen Road, Piscataway, NJ 08854-8019 USA}










\begin{abstract}

The star formation histories (SFHs) of galaxies contain imprints of the physical processes responsible for regulating star formation during galaxy growth and quenching. We improve the Dense Basis SFH reconstruction method of Iyer \& Gawiser (2017), introducing a nonparametric description of the SFH based on the lookback times at which a galaxy assembles certain quantiles of its stellar mass. The method uses Gaussian Processes to create smooth SFHs that are independent of any functional form, with a flexible number of parameters that is adjusted to extract the maximum possible amount of SFH information from the SEDs being fit.  We apply the method to reconstruct the SFHs of 48,791 galaxies with $H<25$ at $0.5 < z < 3.0$ across the five CANDELS fields. 
Using these SFHs, we study the evolution of galaxies as they grow more massive over cosmic time. We quantify the fraction of galaxies that show multiple major episodes of star formation, finding that the median time between two peaks of star formation is $\sim 0.42_{-0.10}^{+0.15}t_{univ}$ Gyr, where $t_{univ}$ is the age of the universe at a given redshift  and remains roughly constant with stellar mass. Correlating SFHs with morphology, we find that studying the median SFHs of galaxies at $0.5<z<1.0$ at the same mass ($10^{10}< M_* < 10^{10.5}M_\odot$) allows us to compare the timescales on which the SFHs decline for different morphological classifications, 
ranging from $0.60^{-0.54}_{+1.54}$ Gyr for galaxies with spiral arms to $2.50^{-1.50}_{+2.25}$ Gyr for spheroids. The Gaussian Process-based SFH description provides a general approach to reconstruct smooth, nonparametric SFH posteriors for galaxies with a flexible number of parameters that can be incorporated into Bayesian SED fitting codes to minimize the bias in estimating physical parameters due to SFH parametrization.









\end{abstract}




\section{Introduction} \label{sec:intro}

Galaxies are massive, turbulent systems, shaped by physical processes that regulate star formation across many orders of magnitude in spatial and temporal scales, \citep[e.g,][]{white1978core, searle1973history, hopkins2014galaxies, genel2018quantification}. Despite this apparent chaos, observations of ensembles of galaxies across cosmic time reveal several correlations, such as the Tully-Fisher relation \citep{tully1977new}, the SFR-M* correlation \citep{noeske2007star, daddi2007multiwavelength, elbaz2007reversal}, 
the black hole mass-velocity dispersion correlation \citep{gebhardt2000relationship}, and the mass-metallicity correlation \citep{tremonti2004origin}. 
Median trends constructed using these scaling relations indicate an equilibrium mode of galaxy growth through baryon cycling, punctuated by mergers and followed by eventual quiescence \citep{peng2010mass, dave2008galaxy, tacchella2016confinement, behroozi2018universemachine}. However, these trends can be equivalently recovered through stochastic evolution \citep{kelson} or simple parametric models with minimal physics \citep{abramson2016return}. Given information about the present state of a galaxy, it is an important question to determine the extent to which its evolution is driven by evolving physical processes such as baryon cycling and star formation suppression, which depend on the physical conditions of galaxies such as their size, morphology and stellar mass; as opposed to stochastic processes governing halo and galaxy mergers and the creation and destruction of molecular clouds that regulate in-situ star formation, which remains broadly invariant across many orders of magnitude as inferred from the SFR-M$_*$ correlation extending from galaxy-wide scales \citep{whitaker2014constraining, kurczynski2016evolution} down to kpc scales in resolved observations \citep{hsieh2017sdss}.

A key observable that correlates the present state of a galaxy with its evolutionary history is its star formation history (SFH) - a record of when a galaxy formed its stars. The SFHs of galaxies bear imprints from all the physical processes that shape galaxy growth by regulating star formation. This includes inflows and outflows of gas, mergers between galaxies, and feedback due to supernovae and Active Galactic Nuclei, which leave imprints on the SFH on timescales ranging from $< 1 Myr$ to $> 10 Gyr$ \citep{somerville2008semi, somerville2015star, sparre2015star, inutsuka2015formation, torrey2017similar, behroozi2018universemachine, matthee2018origin, weinberger2018supermassive}. 

Zeroth order summary statistics of the SFH allow us to calculate traditionally estimated quantities like the stellar masses, star formation rates at the epoch of observation, and mass- and light-weighted ages of individual galaxies \citep{bell2007star}.  First order information about the shape of galaxy SFHs allows us to infer whether the galaxy is actively forming stars, or if it formed most of its stars in the distant past \citep{kauffmann2003stellar, brinchmann2004physical}. Nonparametric estimates of the median SFH for a sample of galaxies let us better understand the width and peak of their median SFHs \cite{moped, vespa, pacifici2016timing, iyer2017reconstruction}, the origin and evolution of scaling relations \citep{iyer2018sfr, torrey2018similar, matthee2018origin}, mass functions (Pacifici+19, in prep.) and the cosmic star formation rate density \citep{leja2018quiescent}.

With the advent of large surveys and the impending arrival of the next generation of surveys with JWST, developing more sophisticated nonparametric techniques of recovering galaxy SFHs will allow us to estimate galaxy properties out to higher redshifts. More importantly, more precise multiwavelength data allows us to estimate \textit{second order} features in galaxy SFHs, ie. fluctuations about the median SFH, seen in the form of SFHs with multiple strong episodes of star formation that can be caused by violent events like mergers or smoother events like stripping followed by inflow of pristine gas \citep{kelson, torrey2018similar, tacchella2018redshift, boogaard2018muse}. For example, \citet{morishita2018massive} find that even old, quiescent galaxies sometimes show evidence for multiple episodes of star formation. Correlating these features of the SFH with other observables such as the metallicity of the galaxy, evidence of recent mergers, and environmental conditions will allow us to test different models that can help explain the diversity seen in SFHs at a particular epoch. 

In the observational domain, the integrated light from distant galaxies contains a host of information about the physical processes that shape them during their formative phases \citep{tinsley1968evolution, bc03,conroy2009propagation, conroy2010propagation}. Since stellar populations of different ages have distinct spectral characteristics, careful analysis of multiwavelength spectral energy distributions (SEDs) allows us in principle to disentangle these different populations \citep{moped, moped2, vespa, dye, acquaviva2011sed, pacifici, smith2015deriving, pacifici2016timing, iyer2017reconstruction, dominguez2016pathways, lee2017intrinsic, ciesla2017sfr, carnall2018measure, leja2018measure}.
SED fitting based SFHs allow us to estimate the SFHs for a much larger population of galaxies, but require much more sophisticated analysis to avoid biases. Assumptions of simple parametric forms for SFHs lead to biases, as shown in \citet{iyer2017reconstruction, ciesla2017sfr, lee2017intrinsic, carnall2018measure}. On the other hand, more complicated parametric forms, as well as methods that estimate the SFR in time bins require us to estimate a much larger number of parameters. Therefore, we are in a situation where we would like to estimate as much information about the SFH as possible, encoded in the smallest possible number of estimated variables, while attempting to be as non-parametric as possible to avoid biases. 

In \citet{iyer2017reconstruction} we introduced the Dense Basis method, which uses a basis of SFHs comprised of four different functional families and all of their combinations, determining the optimal number of SFH components using a statistical test. While this approach produces a basis of SFHs that is effectively dense in SED space and was shown to minimize the bias and scatter due to SFH parametrization, it still retains a minor dependence on the functional families under consideration. Additionally, it can not be flexibly incorporated into a MCMC or nested sampling framework, and becomes computationally expensive as we go to large numbers of SFH components - making it inefficient at extracting all the SFH information present in high quality spectrophotometric data. 

In this work, we introduce an improved version of the Dense Basis method that uses nonparametric SFHs constructed using Gaussian Process Regression, using the lookback times at which a galaxy assembled certain quantiles of its overall stellar mass. Gaussian Processes \citep{gp_book} extend the gaussian probability distribution to the space of functions, allowing us to describe posteriors in SFH space without the need for parametric forms or bins in time. While retaining all the advantages of the previous method, this has the additional advantages of being completely independent of any choice of functional form, scaling linearly with the number of SFH parameters, and providing a modular framework capable of being incorporated into any existing SED fitting routine. We establish the robustness of the method using Semi-Analytic Models and Hydrodynamical Simulations for which the true SFHs are known, and apply it to galaxies at $0.5<z<3.0$ across the five CANDELS fields to study the evolution of galaxies around cosmic noon using their SFHs.

This paper is structured as follows: Section \ref{sec:method} describes the methodology, including the formalism used for constructing smooth, nonparametric SFHs and incorporating them into a SED-fitting framework. In section \ref{sec:data}, we introduce the CANDELS dataset that we use for the current analysis. In section \ref{sec:results}, we describe the SFHs reconstructed from the CANDELS sample, including the fraction of galaxies with multiple strong episodes of star formation, the evolution of this fraction with time, implications for the timescales of quenching followed by rejuvenation as well as for the morphological transformation of galaxies as they approach quiescence. \S\ref{sec:discussion} considers caveats and future directions for applying the Dense Basis method, and \S\ref{sec:conclusions} concludes. 
Appendix \ref{app:validation} contains a suite of validation tests for the methodology developed and applied in this work. Throughout this paper magnitudes are in the AB system; we use a standard $\Lambda$CDM cosmology, with $\Omega_m = 0.3$, $\Omega_\Lambda = 0.7$ and $H_0 = 70$ km Mpc$^{-1}$ s$^{-1}$.

\section{Methodology} \label{sec:method}

\subsection{Star Formation Histories}
\label{sec:sfh_details}

\begin{figure*}
    \centering
    \includegraphics[width=500px]{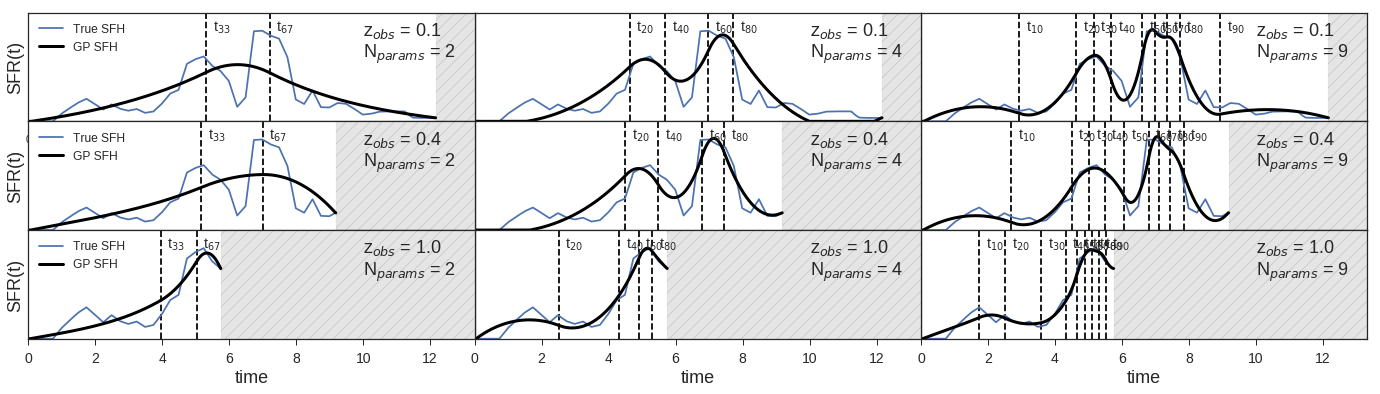}
    \caption{Applying Gaussian Process regression to create smooth approximations of a single SFH from the Santa Cruz semi-analytic model \citep{somerville2008semi, porter2014understanding} at different redshifts of interest using the parametrization described in Sec. \ref{sec:sfh_details}. The nine panels show how versatile the method is at capturing details of the SFH using a varying number of parameters, which can be tuned based on the quality of the observations such as rest-frame wavelength coverage or S/N. In SFH space, this demonstrates the versatility of the method at describing arbitrary SFH shapes, with the accuracy of the approximation increasing with the number of parameters. An advantage of the gaussian process formalism is that it is possible to describe an SFH with multiple episodes with as little as 3 parameters.}
    \label{fig:gp_sfh_example}
\end{figure*}

The main thrust of the Dense Basis method is to encode the maximum amount of information about the SFHs of galaxies using a minimal number of parameters. In this respect, the formalism used to describe SFHs in this work can be used as a module in existing sophisticated inference methods developed in public SED fitting codes like Bagpipes \citep{carnall2018measure}, Beagle \citep{chevallard2016modelling}, CIGALE \citep{noll2009analysis}, and Prospector \citep{leja2017deriving} to flexibly describe SFHs. As shown in our validation tests (appendix \ref{sec:validation}), this description minimizes the bias in estimating SFHs at all lookback times, compared to both existing parametric and nonparametric methods \cite{iyer2017reconstruction, lee2010estimation, ciesla2017sfr,carnall2018measure, leja2018measure}.

This subsection describes the methodology for creating SFHs using this formalism for a given number of parameters, and the following subsections handle the construction of the SED fitting machinery needed to infer the optimal number of SFH parameters corresponding to the amount of information available in individual galaxy SEDs. 

We define a SFH by the tuple: $( M_*, SFR, \{t_X\} )$, where $M_{*}$ is the stellar mass, $SFR$ is the star formation rate at the epoch of observation, and the set $\{ t_X \}$ are N `shape' parameters that describe the  SFH. The shape parameters $\{ t_X\}$ parameters are N lookback times at which the galaxy formed equally spaced quantiles of its total mass \citep{pacifici2016timing, behroozi2018universemachine}. For the first few values of N, we can write
\begin{align*}
    N &= 1 ~~~~ P = \{t_{50} \} \\
    N &= 2 ~~~~ P = \{t_{33}, t_{67} \} \\
    N &= 3 ~~~~ P = \{t_{25}, t_{50}, t_{75} \} \\
    N &= 4 ~~~~ P = \{t_{20}, t_{40}, t_{60}, t_{80} \} \\
    & ...
\end{align*}
This can be seen graphically in Figure \ref{fig:gp_sfh_example}, where we show the $\{ t_x\}$ parameters with $N = 2,4,9$ (vertical dashed black lines) for a mock SFH (blue line), along with our construction of the SFH using these parameters (black solid lines). As expected, the shape of the mock SFH is better captured as N increases, with multiple episodes captured using four parameters. In practice, we find that it is possible to recover multiple episodes of star formation with as little as 3 $\{t_x\}$ parameters. Together with the stellar mass and SFR, this tuple describes a set of integral constraints that describe the shape and overall normalization of the SFH. For a galaxy at redshift z, when the universe was $t_z$ Gyr old, the constraints are:
\begin{align}
    M_{*}(t_z) = \int_{t=0}^{t_z} SFH(t) f_{ret}(t-t_z,Z) dt \nonumber \\ 
    \left \{ \frac{i M_{*,tot}(t_z)}{N+1}  = \int_{t=0}^{t_{x,i}} SFH(t) dt \right \} \forall i \in N
    \label{eqn:constraints}
\end{align}
where M$_*$ is the present stellar mass of the galaxy, M$_{*,tot}$ is the total mass in stars formed during the galaxy's lifetime, and $f_{ret}(t-t',Z)$ is a metallicity dependent fraction of the mass of formed stars that is retained as stars or stellar remnants at the time of observation  \cite{conroy2009propagation, conroy2010propagation}, typically between $0.6 - 1$\footnote{The first equation could alternatively be phrased in terms of $M_{*,tot}$ the total mass in stars formed during the galaxy's lifetime, related to the stellar mass by the equation,
\begin{equation}
    M_* = \frac{\int_{t=0}^{t_z} SFH(t) f_{ret}(t-t_z,Z) dt}{\int_{t=0}^{t_z} SFH(t) dt} M_{*,tot}
\end{equation}}. The second line is a set of N equations, one for each parameter in the set $\{ t_x \}$ that requires that the galaxy form `x' fraction of its total mass by time $t_x$.

This description of a galaxy's SFH already offers several advantages over methods found in the literature, a few of which are summarized below:
\begin{enumerate}
    \item Not being restricted to a particular functional form minimizes bias due to SFH parameterization \citep{iyer2018sfr}.
    \item Describing an SFH using $\{t_X \}$ reduces the discrepancy in S/N per parameter in comparison to methods that determine the SFR in bins of lookback time, since here for example $t_{20}$ might be less well constrained compared to $t_{80}$, but the overall signal depends on the shape of the SFH. e.g. the parametrization will not try to constrain the SFR in the first year after the big bang unless enough stars were formed that early to provide a discernible signal in the SED. This can be compared to methods that adaptively choose time bins eg. VESPA \cite{vespa}.
    \item This provides an ideal framework for compressing the amount of information present in an SFH to a small set of numbers given a way to reconstruct an SFH from a tuple, and hence for comparing SFHs across different simulations and observations on the same footing. 
    \item The distribution of different $\{ t_X \}$ among galaxies at a given epoch within a simulation can be extremely useful in defining and checking the physical assumptions of the SFH priors during SED fitting.
\end{enumerate}

Reconstructing an SFH from the tuple $( M_*, SFR, \{t_X\} )$ can be done in multiple ways, but we seek to minimize the information lost in doing so, while remaining computationally inexpensive. As with all compression methods, we approach the true SFH as the number of parameters N in the set $\{ t_X\} \to \infty$, but we would like to minimize the loss even with a relatively small number of parameters. 

Reconstructing an SFH requires quantifying the integral constraints as points on a fractional mass ($M_{*,tot}(t)$) - cosmic time ($t$) plane and drawing a piecewise smooth curve passing through these points. Rescaling the mass axis and differentiating this cumulative curve would then yield the SFH as a star formation rate at each lookback time. The simplest approximation would be to connect each point such that the resulting SFH is piecewise linear, and while this provides an acceptable solution, it is not very physical in the sense that taking the derivative yields a SFR with jump discontinuities. While generalizing to polynomials for the interpolation causes problems with the derivative going negative in parts of the SFH, methods such as tensioned cubic splines and piecewise-cubic hermite polynomial interpolation (PCHIP) \citep{de1978practical, butt1993preserving} provide more sophisticated solutions to this problem.

In this work, we use Gaussian Process Regression \citep{gp_book, leistedt2016data} implemented through the \verb|george| python package \citep{foreman2015george, hodlr} to create a smooth SFH along with uncertainties following a physically motivated covariance function (kernel) for a given SFH tuple. The Gaussian Process framework uses a set of constraints, given by the set of equations \ref{eqn:constraints}, along with a covariance function (or kernel) to estimate the probability of SFH(t) at a given time t. We use a Matern32 kernel \citep{seeger2004gaussian} in the present application, where the covariance function contains a scale length hyperparameter that sets how much the SFR(t) can vary from the SFR(t$+\Delta t$) separated by a time interval $\Delta t$. The hyperparameter in the kernels essentially encode the amount of stochasticity in the SFHs. This is tested using SFHs from the Santa Cruz semi-analytic models \citep{somerville2008semi, porter2014understanding} and the \citet{dave2016mufasa} MUFASA simulation to minimize loss during the reconstruction of SFHs. In practice, this can be thought of as setting the tension in a string that passes through all the constraints in fractional mass-cosmic time space, and therefore affects the shape and amount of ringing that can happen between two quantiles. While using a spline to reconstruct the SFH, this ringing can sometimes cause the SFR to be negative. Our choice of physically motivated kernel minimizes this behaviour and limits it to pathological SFH tuples (e.g. $t_{25}$ = 0.1 Myr, $t_{50}$ = 10 Gyr, $t_{75}$ = 10.1 Gyr)  which account for $< 3\%$ across our basis. In these cases we set the relevant portion to 0 and check that the overall error in stellar mass due to this is $< 5\%$.  Examples of this approximating SFHs using this formalism are shown in Figure \ref{fig:gp_sfh_example}. The advantage of this method is that both parameters (M$_*$, SFR, $\{t_x\}$) and uncertainties on these parameters can be passed as arguments while constructing the SFH posterior.

\subsection{SED fitting}

\begin{figure*}
    \centering
    \includegraphics[width=500px]{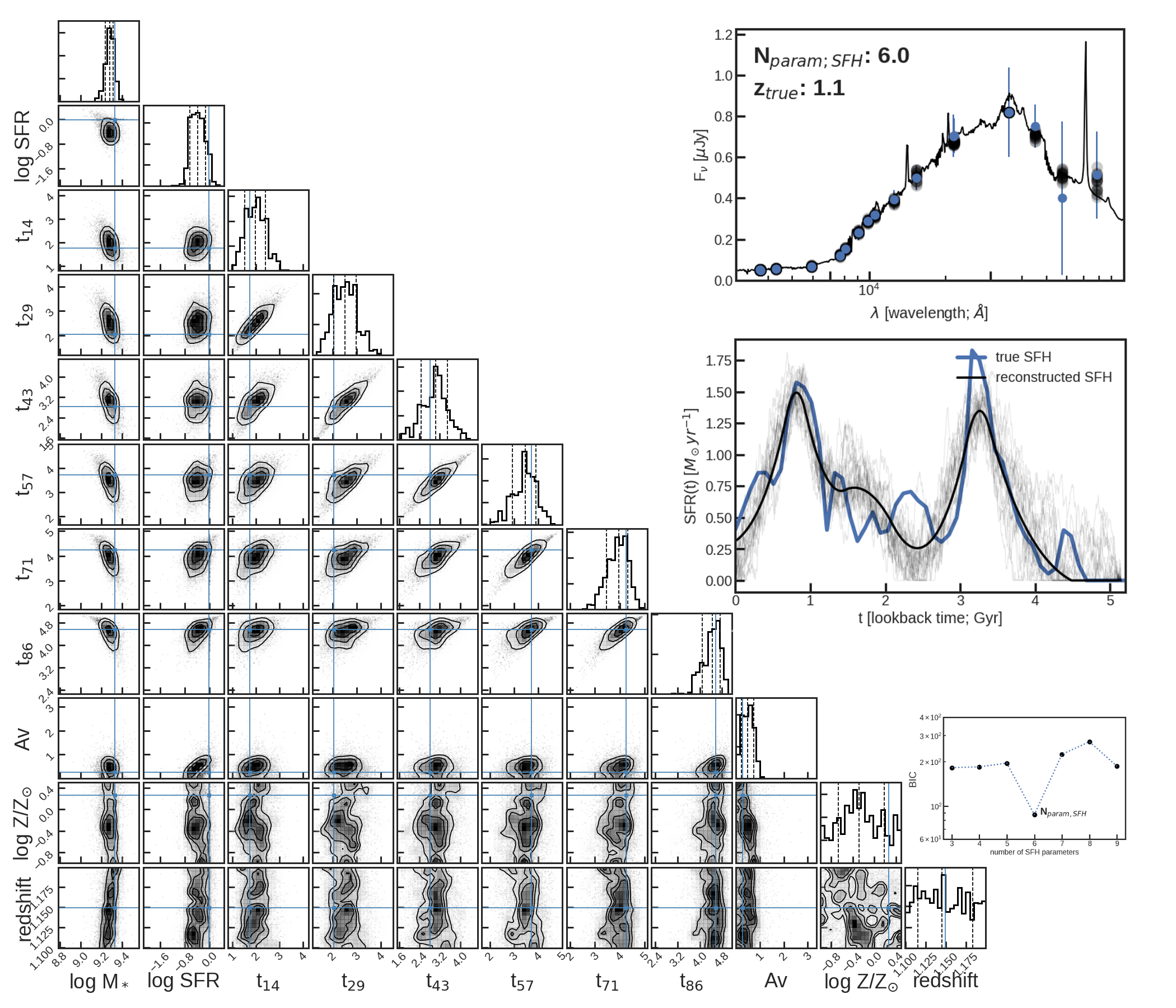}
    \caption{An illustrative example of the SED fitting method, applied to mock noisy photometry for a galaxy from the \citet{somerville2015star} semi-analytic model with more than one major episode of star formation, with the photometry simulated using the GOODS-S filter set. The \textbf{top-right} panel shows the simulated photometry (blue errorbars) and the spectrum (solid black line) corresponding to the median parameter values. The corresponding reconstructed SFH (solid black line) is shown in the \textbf{middle-right} panel below it, with the true SFH (solid blue line) from the SAM shown for comparison. The thin black lines in both panels show draws from the posterior distribution for comparison. This particular SED was determined to contain enough information to estimate 6 correlated SFH parameters using the BIC model selection criterion, as shown in the inset panel (\textbf{bottom right}). The corner plot (\textbf{left}) shows the posteriors for each parameter using our brute-force bayesian approach, with the blue lines representing the true values used to generate the mock noisy photometry. Although redshift is formally a free parameter, we fit galaxies at the ($z_{best} \pm 0.05$) from Kodra et al. (in prep.), finding that within this small dynamic range the redshift posterior is effectively flat, as expected. Dashed black lines for each histogram show the median and 16-84$^{th}$ percentile range.}
    \label{fig:sed_fitting_example}
\end{figure*}

While we are primarily interested in the star formation history, to determine this from the galaxy's SED we need to account for several other factors such as the chemical enrichment, stellar initial mass function (IMF), dust attenuation model, and absorption by the inter-galactic medium (IGM). We then formulate the SFH estimation as an inference problem given by,
\begin{align}
    &P(SFH, A_V, , z ... | F_{\nu,j}^{obs}) = \nonumber  \\ &~~~\frac{P(F_{\nu,j}^{obs} | SFH, A_V, Z, z ... )P(SFH, A_V, Z, z ...)}{P(F_{\nu,j}^{obs})}
\end{align}
The term $P(F_{\nu,j}^{obs} | SFH, A_V, Z, z )$ is the likelihood, given by $max(\mathcal{L }\propto exp(-\chi^2/2))$, where 
\begin{equation}
    \chi^2 = \sum_{j=1}^{N_{filters}} \left( \frac{F_{\nu, j}^{obs} - F_{\nu, j}^{model} (SFH, A_V, Z, z) }{\sigma_j} \right)^2
    \label{eqn:chi2}
\end{equation}
The term $P(SFH, A_V, Z, z ...)$ denotes the prior distribution for the model. If we assume uncorrelated priors for all the parameters, this can be written as $P(SFH)P(A_V)P(Z)P(z)...$. $F_{\nu,j}^{obs}$ is the observed photometry being fit, in the $j^{th}$ photometric filter. $SFH$ denotes the star formation history tuple $( M_*, SFR, \{t_X\} )$, $A_V$ is the dust model, $Z$ the stellar metallicity and z is the redshift. In addition to this, we need to consider the stellar population synthesis models, stellar initial-mass function, absorption by the intergalactic medium, and a self-consistent implementation of nebular emission lines using CLOUDY through FSPS. 
We adopt the Calzetti attenuation law for the dust attenuation \citep{calzetti2001dust} with 1 free parameter since the Calzetti law couples the birth-cloud attenuation to that from older stars and one for stellar metallicity, a Chabrier initial mass function \citep{chabrier2003galactic} with no free parameters, and define our SFH parametrization in section \ref{sec:sfh_details} with N+2 parameters, where N is the number of SFH percentiles, given by the set $\{t_x\}$. We use the Flexible Stellar Population Synthesis \verb|FSPS| code \citep{conroy2009propagation, conroy2010propagation} to generate spectra corresponding to the Basel stellar tracks and the Padova isochrones. With N SFH quantiles, the model then has N + 5 ($M_*, SFR, A_V, Z, z$) free parameters that need to be determined from the data. We construct the method in a way that N itself is a variable that is tuned extract the maximum amount of information present in a galaxy's SED.
It is important to keep in mind that these modeling choices can impose an implicit prior on our SED fits. The effects of testing our model assumptions and priors are further explored in Appendix \ref{sec:validation}, where we find that our models and priors are suitably robust considering the S/N and wavelength coverage of our dataset.

We implement the posterior computation numerically using a brute-force bayesian approach similar to \citet{pacifici, pacifici2016timing, dacunha2008simple} using a large pre-grid of  model SEDs constructed through random draws from the prior distributions corresponding to each free parameter in Eqn. \ref{eqn:chi2}. To ensure that the  pre-grid samples the priors finely enough and is effectively dense in SED space, we perform fits to a sample of 1000 galaxies while varying the size of our pregrid. Using this, we estimate the optimal size of the pregrid as the point where the improvement in median $\chi^2$ for the sample as a function of pregrid size is negligible, leading to a pre-grid with $\sim 900,000$ SEDs. 

\begin{figure}
    \centering
    \includegraphics[width=240px]{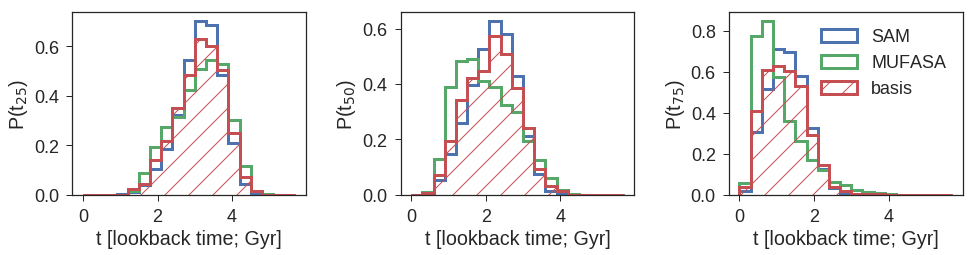}
    \caption{The prior distributions of SFHs in $t_{25}, t_{50},t_{75}$ space for SFHs of galaxies at $z\sim 1$ from the Santa Cruz semi-analytic model, and the MUFASA hydrodynamical simulation, in addition to the basis assembled using the Dirichlet prior we adopt.}
    \label{fig:sfh_prior_dists}
\end{figure}

To construct the pre-grid, we draw random values from our prior distributions for stellar metallicity, dust attenuation, and SFH parameters ($M_*, SFR, \{t_x\}$).
For metallicity, we adopt a flat prior on $\log Z/Z_\odot$ \citep{pacifici2016timing, carnall2018measure, leja2018measure}, an exponential prior on dust attenuation \citep{iyer2017reconstruction}, and a Dirichlet prior for the lookback times $\{ t_x\}$ that specify the shape of the SFH. The Dirichlet prior is a generalization of the Beta distribution to N variables, such that a random draw yields N random numbers $x_i$ that satisfy $\sum_i x_i = 1$. More details about this prior can be found in \citet{leja2017deriving, leja2018measure}. In practice, for a galaxy SFH at redshift z, we generate the set $\{ t_x\}$ by performing a random draw multiplied by the age of the universe at that redshift, giving the set of lookback times at which a galaxy formed various quantiles of its stellar mass. The dirichlet prior has a single tunable parameter $\alpha$ that specifies how correlated the values are. In our case, values of the concentration parameter $\alpha < 1$ results in values that can be arbitrarily close, leading to extremely spiky SFHs since galaxies have to assemble a significant fraction of their mass in a very short period of time, and $\alpha > 1$ leads to smoother SFHs, with more evenly spaced values that nevertheless have considerable diversity. In practice, we use a value of $\alpha = 5$, which leads to a distribution of parameters that is similar to what we find in the SAM and MUFASA. 
This can be seen in Figure \ref{fig:sfh_prior_dists}. 

SFH uncertainties are computed after the fit is performed, using the posterior for each parameter describing the tuple $( M_*, SFR, \{ t_x \} )$. For each galaxy, 100 self-consistent random draws are performed from the posterior with the covariances between parameters taken into account, and corresponding SFH realizations are constructed using the Gaussian Process routine. For a given realization, if the set of parameters already exist in the pre-grid, the corresponding precomputed SFH is used to decrease the computational cost. Figure \ref{fig:sed_fitting_example} shows 20 draws from the SFH posterior for that galaxy using this approach as thin black lines in the SFH inset panel. 68$\%$ confidence intervals are then constructed by taking the $16^{th}$ to $84^{th}$ percentile of the SFR distribution at each point in lookback time.

\subsection{Nonparametric SFH reconstruction}

The Gaussian Process based SFH (GP-SFH) formalism allows us to gain independence from having to make a choice of functional form for the shape of the SFH, without having to bin SFHs in lookback time. This results in smooth, effectively nonparametric SFHs that minimize the bias and scatter in SFH reconstruction, as seen in Sec. \ref{sec:validation}. However, since we would like the method to be truly nonparametric, we require that the number of $\{ t_x\}$ parameters be optimized for the amount of information present in a given noisy SED. To implement this in practice, we generate pre-grids for SFHs with the set $\{t_x\}$ ranging from 3 to 9 parameters, since it is difficult to specify the shape of complex SFHs with less than three parameters, and it is impractical to recover more than nine from broadband SEDs. We then fit each observed SED with all 7 pre-grids, and obtain the most appropriate number of parameters using an appropriate model selection criterion. Ideally the bayesian evidence would be used for this model selection step \citep{liddle2007information}. However, in practice, the numerical computation of the evidence is expensive due to the need for a nested sampler, and can not be completed for the number of galaxies typically found in large photometric surveys. Having tested the properties of the likelihood surfaces using this method, however, we find that while they may be multimodal, they generally do not contain pathological features that necessitate this numerically expensive procedure. In light of this, we perform our model selection using an approximation of the evidence, given by the Bayesian Information Criterion \citep{schwarz1978estimating,liddle2007information}, defined as 
\begin{equation}
    BIC = k \ln (n) - 2 \ln \mathcal{L}_{max}
\end{equation}
where $n$ is the number of photometric datapoints, $k$ is the number of parameters in the SFH tuple, and $\mathcal{L}$ is the maximized value of the likelihood function given by Eqn. \ref{eqn:chi2}. The latter term in this equation is a measure of how well the model describes the data, and the former term is a penalty for an increased number of parameters. While the BIC does not account for degeneracies between parameters, the effect of this leads an exaggerated effect of the parameter penalty term, leading to more conservative estimates of N$_{param}$. By finding the minima in the BIC, we find the number of SFH parameters that can be robustly extracted from the SED being fit. This leads to a truly nonparametric description of the SFH, based on the amount of information about the different stellar populations encoded in the galaxy's SED. An example of this for a single galaxy is shown in the bottom right plot of Figure \ref{fig:sed_fitting_example}. 
\citet{iyer2017reconstruction} used a similar nonparametric estimate of the number of SFH components using a F-test, but was computationally expensive to implement as the number of SFH components grew. \citet{vespa} and \citet{dye} also used nonparametric methods to estimate the optimal number of time bins during SFH reconstruction, although this is prone to bin edge effects, as described in \citet{leja2018measure}.

\subsection{Quantifying the number of Major Episodes of Star Formation in a SFH}

Unlike simple SFH parametrizations commonly used in SED fitting, the Gaussian Process based star formation histories can have multiple maxima, even with four or fewer $t_X$ parameters, as seen in Figure. \ref{fig:gp_sfh_example}. 
Hence, it is possible to analyze them using a peak finding algorithm to quantify the number of major episodes of star formation in a galaxy's past. For this particular analysis, we use the best-fit SFH for each galaxy, since the median SFH for each galaxy is biased towards smooth SFHs. 
This is an effect due to our choice of kernel, which prefers smooth solutions when the $\{t_x\}$ parameters are uncertain, as seen in the left column of Figure \ref{fig:gp_sfh_example}. 
With tighter constraints on $\{ t_x\}$ from higher S/N data this problem is alleviated. As a result, while the number of major episodes ($N_{ep}$) estimated using this method for individual galaxies is susceptible to noise, the overall distribution is seen to be recovered without any significant bias, as shown in Appendix \ref{sec:validation}. 

For each SFH, we quantify the number of episodes as follows: We first find the number of peaks in an SFH as the set of points that satisfy $d SFH/dt = 0$ and $d^2SFH/dt^2 < 0$. To separate multiple peaks within an overall episode of star formation from different episodes, we impose a peak prominence criterion by requiring that
\begin{equation}
(\log SFR_{peak} -  \log SFR_{min,local}) > 1.5 + \frac{1.5}{4}\log \frac{M_*}{10^8M_\odot} 
\label{eqn:sfr_threshold}
\end{equation}
where $SFR_{min,local}$ is the minima between two peaks in the SFH. This condition is shown in Appendix \ref{sec:validation} to minimize the type-1 (overestimating $N_{ep}$) and type-2 (underestimating $N_{ep}$) errors in our validation sample. It arises because the sensitivity to star formation drops approximately logarithmically with time \citep{ocvirk2006steckmap}. 
Since more massive galaxies tend to have older stellar populations, we found that a mass-independent peak-prominence criterion caused a mass-dependent bias in our estimates of the number of episodes. We correct for this effect by requiring a more (less) stringent dip in the SFH for a massive (low-mass) galaxy, and while this does not improve the result for every galaxy, it accurately recovers the distribution, which is the quantity that we are interested in. 

\subsection{Validation} 
\label{sec:validation}

At each state of the method development, we performed validation using an ensemble of galaxies from the Santa Cruz semi-analytic models \citep{somerville2008semi, porter2014understanding} and the MUFASA hydrodynamical simulation \citep{dave2016mufasa}. For these tests, we created a mock catalog with 10,000 galaxies at $0.5<z<3.0$, matching our analysis sample. 

For each galaxy in our mock catalog, we draw a random redshift $z_{mock}$ beween $0.5$ and $3$. We then create synthetic spectra using FSPS with the corresponding star formation history and mass-weighted metallicity, with dust attenuation sampled from an exponential distribution as in \citet{iyer2017reconstruction}. We then multiply these spectra by the appropriate filter transmission curves corresponding to one of the five CANDELS fields and perturb the photometry in each band by adding realistic noise derived from the median photometric uncertainties for the CANDELS catalog in the redshift range $[z_{mock}-0.1, z_{mock}+0.1]$. 

Figure \ref{fig:sed_fitting_example} shows an example following this procedure, using a galaxy with a SFH that is not well approximated by a simple parametric form currently used in the literature, but is recovered well with the GP-SFH approach. Using our mock catalog, we then perform a series of validation experiments divided into four categories. In Appendix \ref{app:SFH_validation} we consider the robustness of the reconstructed SFHs, and in Appendix \ref{app:uncert_validation} we consider the robustness of the uncertainties on the reconstructed SFHs. In Appendix \ref{app:param_validation} we quantify the bias and scatter in estimating stellar masses, star formation rates, dust attenuation and stellar metallicities. In Appendix \ref{app:nep_validation} we test the recovery of the number of major episodes of star formation for our mock sample, find a mass-dependent bias and correct for it.

\section{Data} \label{sec:data}

\begin{table*}[ht!]
    \begin{tabular}{c|c}
    \hline \hline
    Field  & Filter set \\
    \hline
    GOODS-S \citep{guo2013candels} & Blanco/CTIO U, VLT/VIMOS U, \\
    & HST/ACS f435w, f606w, f775w, f814w, f850lp, \\
    & HST/WFC3 f098m, f105w, f125w, f160w, \\
    & VLT/ISAAC Ks, VLT/Hawk-I Ks, \\
    & Spitzer/Irac 3.6$\mu$m,  4.5$\mu$m,  5.8$\mu$m,  8.0$\mu$m   \\
    GOODS-N (Barro et al., in prep.) & KPNO U, LBC U, \\
    & HST/ACS f435w, f606w, f775w, f814w, f850lp, \\
    & HST/WFC3 f105w, f125w, f140w, f160w, f275w, \\
    & MOIRCS K, CFHT Ks, \\
    & Spitzer/Irac 3.6$\mu$m,  4.5$\mu$m,  5.8$\mu$m,  8.0$\mu$m \\
    UDS \citep{galametz2013candels} & CFHT/MegaCam u, Subaru/Suprime-Cam B, V, Rc, i', z', \\
    & HST/ACS f606w, f814w, HST/WFC3 f125w, f160w, \\
    & VLT/Hawk-I Y, Ks, \\
    & WFCAM/UKIRT J, H, K, \\
    & Spitzer/Irac 3.6$\mu$m,  4.5$\mu$m,  5.8$\mu$m,  8.0$\mu$m \\
    EGS \citep{stefanon2017candels} & CFHT/MegaCam U*, g', r', i', z', \\
    & HST/ACS f606w, f814w, HST/WFC3 f125w, f140w, f160w, \\
    & Mayall/NEWFIRM J1, J2, J3, H1, H2, K, \\
    & CFHT/WIRCAM J, H, Ks,  \\
    & Spitzer/Irac 3.6$\mu$m,  4.5$\mu$m,  5.8$\mu$m,  8.0$\mu$m \\
    COSMOS \citep{nayyeri2017candels}  & CFHT/MegaCam u*, g*, r*, i*, z*, \\
    & Subaru/Suprime-Cam B, g$^+$, V, r$^+$, i$^+$, z$^+$, \\
    & HST/ACS f606w, f814w, HST/WFC3 f125w, f160w, \\
    & Subaru/Suprime-cam IA484, IA527, IA624, IA679, IA738, IA767, IB427,  \\
    & IB464, IB505, IB574, IB709, IB827, NB711, NB816, \\
    & VLT/VISTA Y, J, H, Ks, Mayall/NEWFIRM J1, J2, J3, H1, H2, K,  \\
    & Spitzer/Irac 3.6$\mu$m,  4.5$\mu$m,  5.8$\mu$m,  8.0$\mu$m \\
    \hline
    \end{tabular}
    \caption{Collected measurements comprising the UV-to-NIR SEDs of galaxies across the five CANDELS fields}
    \label{table:filtersets}
\end{table*}

\begin{figure*}
    \centering
    \includegraphics[width=500px]{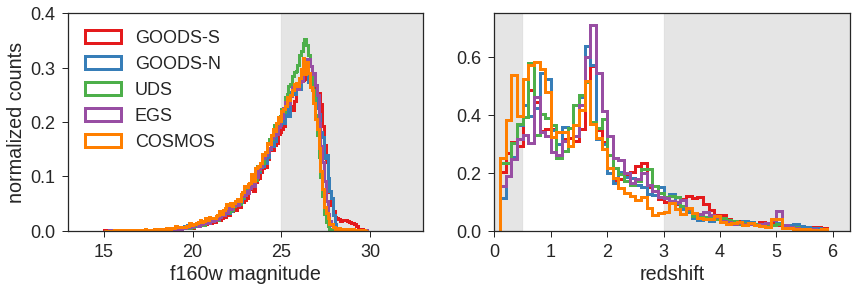}
    \caption{The distribution of galaxies across the five CANDELS fields in HST/WFC3 F160w magnitude (left) and redshift (right). While most redshifts are photometric, the sample contains $\approx$ 7,000 galaxies with spectroscopic redshifts. Grey regions show parts of the sample that we exclude in the current analysis.}
    \label{fig:sample_with_cuts}
\end{figure*}

In the current analysis we use a sample of galaxies from the HST/F160w selected catalogs for the five Cosmic Assembly Near-infrared Deep Extragalactic Legacy Survey (CANDELS) \citep{candels, koekemoer2011candels} fields covering a total area of $\sim 800$ arcmin$^2$ : GOODS-S \citep{guo2013candels}, GOODS-N (Barro et al., in prep.), COSMOS \citep{nayyeri2017candels}, EGS \cite{stefanon2017candels} and UDS \citep{galametz2013candels}.

The GOODS-South \citep{guo2013candels} field contains 34,930 objects and covers an area of $\sim 170$ arcmin$^2$, with a $5\sigma$ limiting depth of 27.4, 28.2, and 29.7 AB magnitudes in the three overlapping survey regions (CANDELS wide, deep, and HUDF regions). The GOODS-North field (Barro et al., in prep.) contains 35,445 objects over a similar area, with a $5\sigma$ limiting depth of 27.5 AB mag \citep{pacifici2016timing}.  The UKIRT Infrared Deep Sky Survey (UKIDSS) Ultra-Deep Survey (UDS) catalog \citep{galametz2013candels} contains 35,932 sources over an area of 201.7 arcmin$^2$ with a $5\sigma$ limiting depth of 27.45 AB magnitudes. The Extended Groth Strip (EGS) catalog \citep{stefanon2017candels} contains 41,457 objecst over an area of $\approx 206$ arcmin$^2$ reaching a depth of 26.62 AB mag. The COSMOS field \cite{nayyeri2017candels} contains 38,671 objects covering an area of $\approx 216$ arcmin$^2$ with a limiting depth of 27.6 AB mag. The catalogs select objects via SExtractor in dual-image mode using F160w as the detection band. The dual image mode \citep{galametz2013candels} is optimized to detect both faint, small galaxies in `hot' mode without over de-blending large, resolved galaxies detected in `cold' mode. The HST (ACS and WFC3) bands were point spread function (PSF) matched to measure photometry, and TFIT \citep{laidler2007tfit} was used to measure the photometry of ground based and IRAC bands using the HST WFC3 imaging as a template. The SEDs in the five fields include a wide range of UV-to-NIR measurements, with the flux measured in 17, 18, 19, 23, and 43 photometric bands in GOODS-S, GOODS-N, UDS, EGS and COSMOS respectively. The photometric filters used for each field are detailed in Table \ref{table:filtersets}.

In this paper, we focus on galaxies at $0.5<z<3.0$, since we do not have multiple reliable rest-UV measurements from the HST bands at $z < 0.45$, leading to larger uncertainties in SFR and can not accurately constrain the $1.6\mu$m bump at redshifts $z > 3.06$ which is important for robust estimation of stellar masses. 
In addition to estimating robust stellar masses and star formation rates, it is necessary to ensure robust S/N in the rest-optical portion of the SED, which is most sensitive to variations in the SFH. As a proxy to the total S/N, we restrict our sample to galaxies with $H < 25$, where H is the HST/WFC3 F160w band.  After implementing these selection cuts, we are left with a total of 48,791 galaxies. The effect of each selection effect and the total number of galaxies used for the analysis is given in Table \ref{table:sample_selection_numbers} and Figure \ref{fig:sample_with_cuts}. To perform our fits, we use an updated CANDELS photometric redshift catalog by Kodra et al. (in prep.) containing an increased number of spectroscopic redshift measurements as well as photometric redshifts with Bayesian combined uncertainties estimated by comparing the redshift probability distributions of four different SED fitting methods.
We perform our fits using their $z_{\rm best}$ binned to the resolution of our pre-grid, with $\delta z = 0.01$.

\begin{table*}[ht!]
    \begin{tabular}{c|c c c c c | c || c}
    \hline \hline
    Field  & All Objects & Good Flags$^{a}$ & $0.5<z<3$ &  $H< 25^{b}$ & $\chi^2_{red} < 10$ & Final sample & $z_{spec}$ $^c$\\
    \hline
    GOODS-S \citep{guo2013candels} &  34,930 & 31,273 & 22,713 & 8,520 & 8,299 & 8,299 & 1,758\\
    GOODS-N (Barro et al., in prep.) & 35,445 & 34,693 & 26,838 & 9,551 & 9,206 & 9,206 & 2,399\\
    UDS \citep{galametz2013candels} & 35,932 & 26,917 & 21,263 & 10,234 & 10,176 & 10,176 & 538\\
    EGS \citep{stefanon2017candels} & 41,457 & 31,714 & 24,444 & 10,554 & 10,261 & 10,261 & 1,671\\
    COSMOS \citep{nayyeri2017candels}  & 38,671 & 30,070 & 22,092 & 10,883 & 10,849 & 10,849 & 705\\
    \hline
    \end{tabular}
    \caption{The number of galaxies used in our analysis, and the effect of each step in the selection process.\\$^a$: Galaxies with flags indicating no contamination by nearby objects, halos or star spikes, as well as objects with a low stellarity classification given by the SExtractor CLASS STAR output. \\$^{b}$: We select galaxies brighter than 25 mag in the HST/WFC3 F160W band, to ensure enough S/N in the rest-optical regime of the SED necessary for accurate SFH reconstruction.\\$^{c}$: This column gives the number of galaxies with confirmed spectroscopic redshifts in each field for our final analysis sample.}
    \label{table:sample_selection_numbers}
\end{table*}

\section{Results} \label{sec:results}

\subsection{The star formation histories of galaxies at $0.5<z<3$}

\begin{figure}
    \centering
    \includegraphics[width=240px]{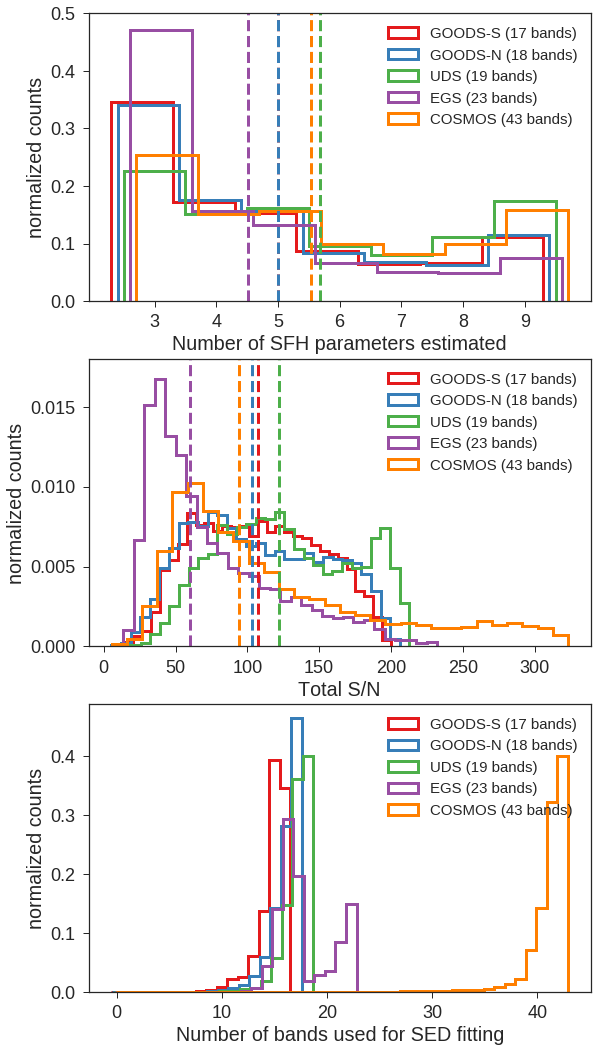}
    \caption{\textbf{Top:} Distributions of the number of SFH parameters estimated while fitting the SEDs of galaxies at $0.5<z<3.0$ in the five CANDELS fields, with the vertical dashed lines showing the mean 
    value of the distribution.
    \textbf{Middle:} The distribution of total S/N $= \sqrt{\sum_j (F_{\nu,j} / \sigma_{\nu,j})^2}$ for the five fields, with the vertical dashed lines showing the medians of each distribution. 
    \textbf{Bottom:} The distribution of the number of photometric bands used in fitting the SEDs in each field. The bimodality in the EGS observations is due to partial coverage of the field with the six NEWFIRM bands (J1,J2,J3,H1,H2,K).}
  \label{fig:num_params_all_fields}
\end{figure}

\begin{figure*}
    \centering
    \includegraphics[width=500px]{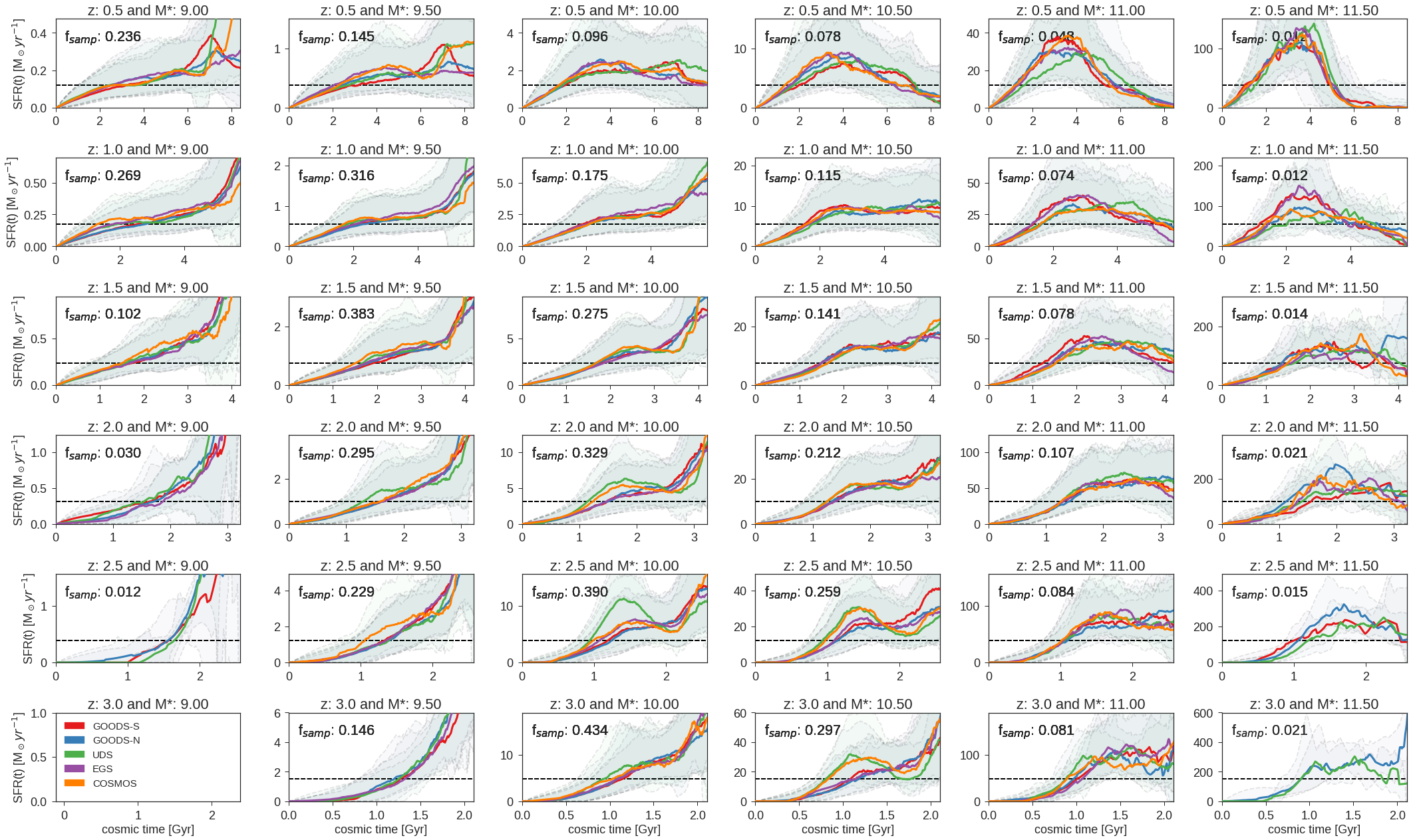}
    \caption{The median star formation histories (SFHs) of galaxies in the five CANDELS fields in bins of stellar mass (horizontal) and redshift (vertical), showing how galaxies evolve with cosmic time and as they grow in stellar mass. Solid colored lines show the median SFH for each field separately, showing a remarkable similarity across the different fields in the majority of the bins. The shaded regions show the $16^{th}-84^{th}$ percentile in the SFHs, highlighting the diversity in SFHs as a function of stellar mass and epoch. The dashed black line shows the mean SFR ($ \equiv M_*/t_{univ}$) assuming constant SFR for that redshift and mass bin, and the $f_{samp}$ at the top-left corner of each panel shows the fraction of all galaxies in our sample at that redshift that fall in a particular mass bin.  In good agreement with cosmological simulations, semi-analytical and empirical models, galaxy SFHs tend to rise with time at low masses and high redshifts, starting to turn over at high masses, with the turnover mass decreasing as we go to lower redshifts. However, in addition to the average SFH behaviour, we also have access to the individual SFH for each galaxy, which now opens up the possibility of repeating this analysis to trace the evolution of SFHs with quantities like $t_{50}$, metallicity, morphology, central density, size, environment, and other probes of galaxy evolution.}
  \label{fig:sfh_all_mass_z}
\end{figure*}

We now apply the Dense Basis SED fitting method as described in Sec. \ref{sec:method} to the sample of CANDELS galaxies at $0.5<z<3.0$ to reconstruct the SFH of each galaxy with uncertainties. Figure \ref{fig:num_params_all_fields} shows the distribution of the number of SFH parameters that individual galaxies are best fit with in each of the five fields, with a slight trend towards increasing amounts of SFH information recovered with more photometric bands.

In Figure \ref{fig:sfh_all_mass_z} we show the distributions of reconstructed SFHs in the five CANDELS fields and how they evolve with mass and redshift. The median and interquartile range are computed at each point in time as in \citet{pacifici2016timing}, who performed a similar calculation for quiescent galaxies using a basis of SFHs derived from a semi-analytic model \citep{de2007hierarchical}. In the current analysis, we have used a redshift bin width of $\delta z = \pm 0.1$, and a mass bin width of $\delta M_* =  \pm 0.25 dex$, such that the $M_* \sim 10^{10}M_\odot$ bin includes all galaxies with $10^{9.75} < M_* < 10^{10.25}M_\odot$. Dashed black lines in each panel show the mean SFR ($ \equiv M_*/t_{univ}$) assuming constant SFR for that redshift and mass bin, and the $f_{samp}$ at the top-left corner of each panel shows the fraction of all galaxies in our sample at that redshift that fall in a particular mass bin. Since there can be a few galaxies that fall outside the plotted mass range, this may not sum to 1.

We see that the median SFH across the five CANDELS fields are remarkably similar, as expected. There are discrepancies in a few of the bins, most notably at $2<z<3$ and $M_*\sim 10^{10}M_\odot$ for UDS, which could be the result of correlated photometric noise, or smearing effects in the photo-z that was used for the calculation. 
In the highest redshift bin, the portion of the SED with wavelengths greater than rest-frame 1.6$\mu m$ are not well sampled, leading to poorer constraints on the SFHs in the the last row. 

In general, SFHs tend to rise at high redshifts and low stellar masses, similar to those from cosmological simulations, with a turnover and subsequent decline as we go to higher masses and lower redshifts. In agreement with \citet{pacifici, behroozi2018universemachine}, massive galaxies tend to peak earlier in their SFHs, and galaxies on average tend to move towards quiescence at lower masses as the universe grows older, with the threshold changing from nearly $10^{11.5}M_\odot$ at $z\sim 3$ to $10^{10}M_\odot$ at $z\sim 0.5$, without the need for any implicit assumptions about the SFHs, such as \citet{behroozi2018universemachine}'s well motivated assumption that earlier forming halos get lower SFRs. The additional advantage of the Dense Basis method is that in addition to the average SFHs, the individual SFHs of galaxies reconstructed from the observations allow us to explore the additional factors that drive the diversity of SFHs at a given mass and epoch. This allows us to extend our analysis beyond the physics encoded in the stellar mass-halo mass relation, which gives a constraint upon the first order behaviour of SFHs.

\subsection{The number of major episodes of star formation experienced by galaxies}

Most studies of galaxy SFHs focus on the overall rise and fall of an ensemble of SFHs \citep{leitnerMSI, pacifici2016timing, ciesla2017sfr, leja2017deriving}, which has led to a well-constrained understanding of the overall behaviour seen in Figure \ref{fig:sfh_all_mass_z}. However, with smooth, nonparametric SFHs it is now possible to ask questions about the \textit{second order} statistics of an ensemble of SFHs, analyzing the departures from this overall behaviour in the form of periods of relative quiescence between episodes of star formation. 

\begin{figure}
    \centering
    \includegraphics[width=180px]{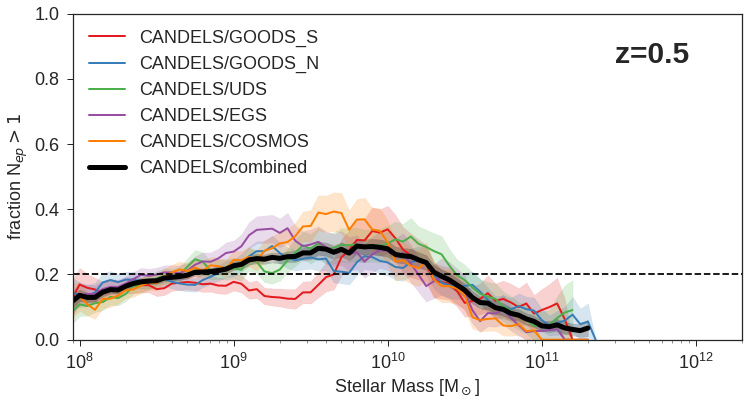}
    \includegraphics[width=180px]{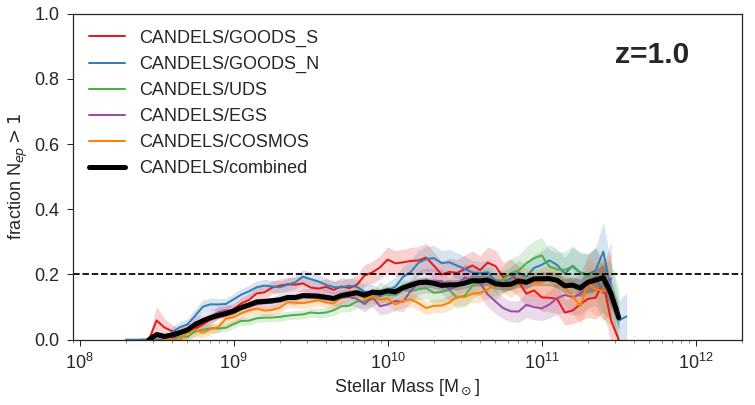}
    \includegraphics[width=180px]{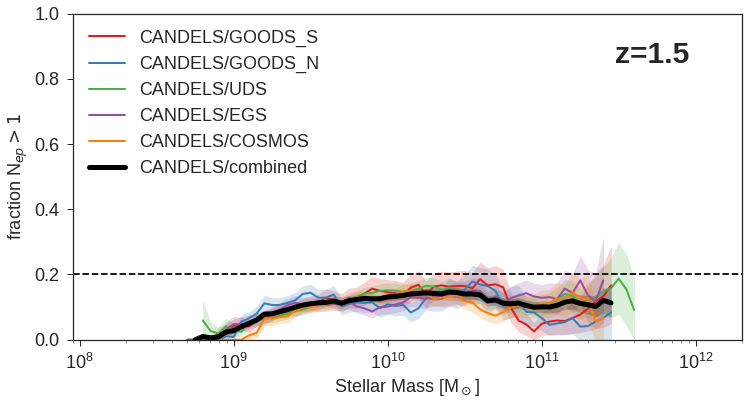}
    \includegraphics[width=180px]{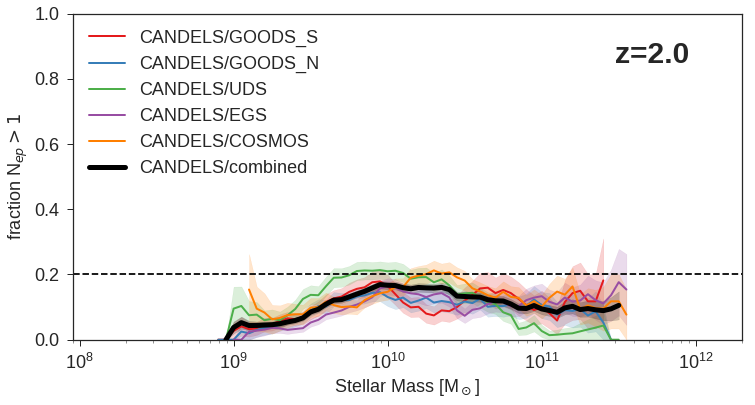}
    \includegraphics[width=180px]{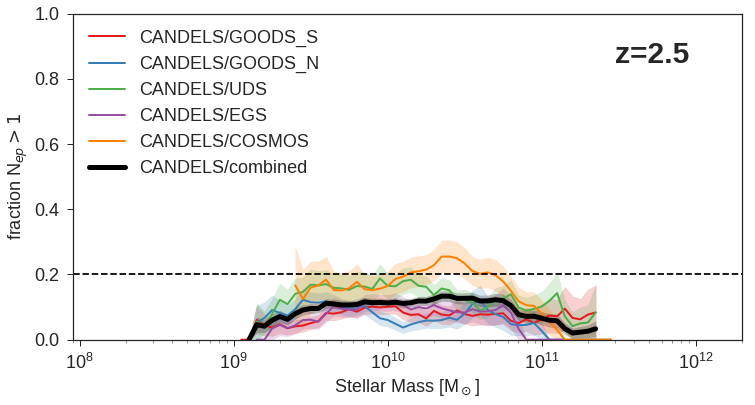}
    \includegraphics[width=180px]{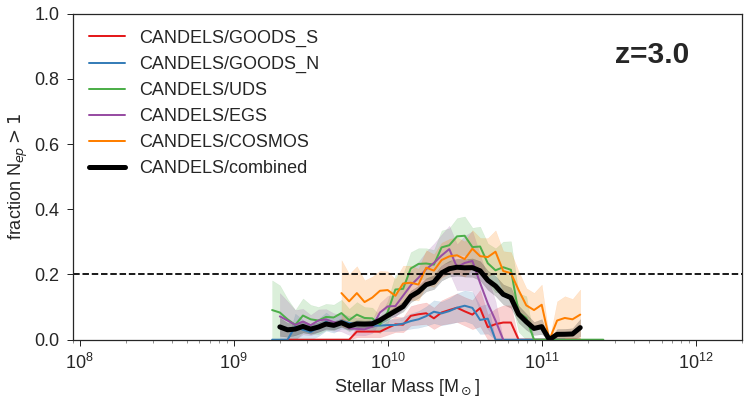}
    \caption{The fraction of galaxies which show multiple major episodes of star formation in their SFHs as a function of stellar mass at various redshifts for galaxies in the five  CANDELS fields. The solid line shows a running median within a bin of $\pm 0.25$ dex in stellar mass, and the shaded regions show the uncertainty for the estimates assuming Poisson noise.}
    \label{fig:nep_redshift}
\end{figure}

In Figure \ref{fig:nep_redshift} we show the fraction of galaxies with different numbers of major episodes of star formation in a galaxy's past at redshifts $0.5<z<3.0$. Since the number of episodes are a discrete quantity, Poisson noise dominates the formal uncertainties in an individual galaxy's number of episodes while calculating functions of the sample, as discussed in Appendix \ref{sec:validation}. The different fields (colored lines), are in good agreement with each other and the median of the full sample (solid black line).

We find that at low redshifts, the fraction of galaxies with multiple major episodes of star formation decreases as we go up in stellar mass above $10^{10.5}M_\odot$, in agreement with \citet{iyer2017reconstruction}.
In addition to this, we find a slight decrease in the overall fraction of galaxies with multiple episodes with increasing redshift at any given mass, with the notable exception of $M_*\approx 10^{10.5}M_\odot$, with does not show a noticeable evolution with redshift. Although we have accounted for S/N variations, the decrease at lower masses could be at least partially due to insufficient S/N to resolve multiple episodes of star formation as we go to lower masses and higher redshifts. 
A few explanations are possible for the behavior at high masses: AGN feedback quenching galaxies \citep{weinberger2018supermassive} could lead to SFHs that form most of their stars at by $z\sim 3$, which could look like a single early episode of star formation without the S/N in the SEDs to resolve the older populations at $z\sim 1$. This is made more probable by the fact that while most galaxies with multiple episodes are found to lie on the SFR-M$_*$ correlation, the greatest number of galaxies with low SFRs at the time of observation and multiple episodes occurs at masses close to $10^{10.5}M_\odot$. Another reason could be the central limit theorem \citep{kelson}: massive galaxies that grow primarily through mergers \citep{brinchmann2004physical, perez2008stellar, bundy2005mass} at early times could be composed of multiple progenitors. As the number of progenitors grows with mass, by the central limit theorem their star formation histories should look smoother than those for less massive galaxies. Low mass galaxies in comparison should have more stochastic star formation histories since they are growing most of their mass in-situ, which would be in close agreement with the findings of \citet{guo2016bursty, matthee2018origin, emami2018closer, shivaei2015investigating} and Broussard et al. (2018), in prep. 

\begin{figure}
    \centering
    \includegraphics[width=240px]{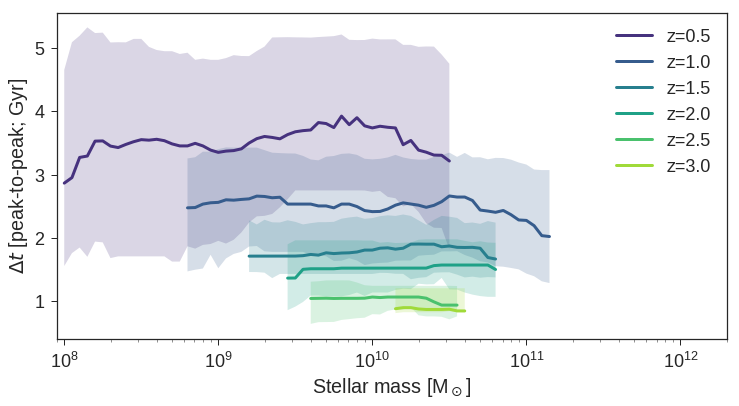}
    \includegraphics[width=240px]{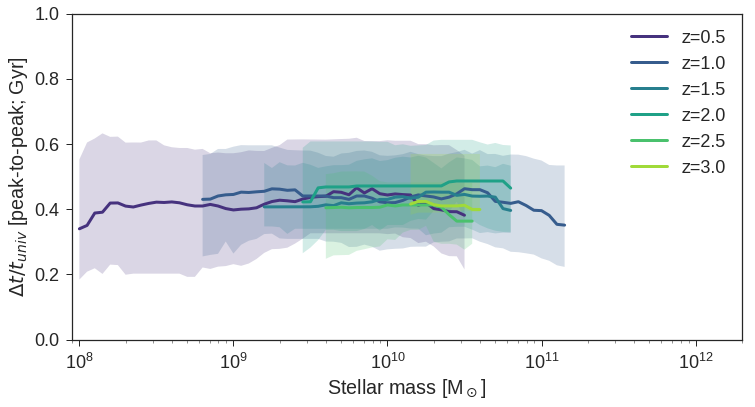}
    \caption{The separation between multiple peaks of star formation, as a function of mass and redshift for the subsample of galaxies that have $N_{ep} > 1$. The redshift bins are the same as fig \ref{fig:nep_redshift} and the solid line and shaded region show the median and 16-84$^{th}$ percentiles respectively. The top panel shows the distribution across stellar mass at different redshifts. The bottom shows the same, but divided by the age of the universe at that epoch.}
    \label{fig:peak_separation_mass_redshift}
\end{figure}

If a galaxy is found to have multiple strong episodes of star formation in its lifetime, an interesting question would be whether the galaxy was actively forming stars and the star formation was temporarily suppressed by a quenching attempt (short time interval between peaks), as opposed to a galaxy that was on its way to quiescence but restarted star formation due to an inflow of pristine gas or merger. This is especially interesting within the context of rejuvenation of galaxy SFRs \citep{fang2012slow}, since it sets timescales for how long a galaxy spends off the star-forming sequence when it makes such an excursion. To quantify this, we measure the time interval between multiple peaks for the subsample of galaxies with $N_{ep} > 1$. We plot this as a function of redshift and mass in Figure \ref{fig:peak_separation_mass_redshift}, finding that although the separation doesn't vary strongly with mass, it does show a strong trend with redshift. However, upon normalizing by the age of the universe at different redshifts, this trend is significantly decreased, leaving us with a roughly constant timescale across which galaxy rejuvenation occurs, given by $t_{\Delta peak} \sim 0.42_{-0.10}^{+0.15} t_{univ}$ Gyr, where $t_{univ}$ is the age of the universe at the redshift of observation. This is similar to the result by \citet{abramson2016return}, which found the transit time through the green valley (i.e. half the time between two peaks for a rejuvenating SFH) to be $\sim 0.2 t_{univ}$ Gyr roughly independently of redshift, and related to the result by \citet{pacifici2016timing} that found that the width of the SFH for quiescent galaxies is roughly constant across stellar mass and redshift when the age of the universe is factored out, and consistent with \citet{muzzin2014phase}, who
find that the post-starburst spectra of galaxies at $z \sim 1$ are well fit with a quenching timescale of $0.4_{-0.3}^{+0.4}$ Gyr. 
\citet{fang2012slow}
identify a subsample of galaxies at $z\sim 0.1$ that could linger in the green valley for $\mathcal{O}$(Gyr). 
The astute reader may wonder how significant it is to find that two episodes are typically separated by roughly half the age of the universe at the time of observation, as this also corresponds to the median period of a generic sine wave possessing two peaks without that interval. Given the present data quality, it is difficult to test this further by investigating the separation between the earliest two star formation episodes in galaxies whose SFHs show 3-5 major episodes of star formation, but that should be done with higher S/N spectrophotometry. At present, we can compare the fit for separation between episodes against the predictions of galaxy formation models, finding similar trends albeit with a slightly smaller value of $\approx 0.3 \pm 0.15 t_{univ}$ Gyr. The consistency of our result with a variety of similar results across a range of redshifts summarized above increases is an additional reassuring check. 
In \citet{behroozi2018universemachine}, galaxy rejuvenation is a generic feature of a population, depending on the mode by which a halo is accreting mass: through mergers, accretion from another halo or infall. In this scenario, rejuvenation can occur more often when the quenched population evolves more slowly than the halo dynamical time, during which it can switch between modes of accretion, increasing or decreasing the SFR as a result. However, our result seems to indicate that the rejuvenation timescales remain relatively constant over cosmic time. It is important to note that the scatter in this quantity is quite large, and the evolution in the quenched fraction happens most rapidly at redshifts $0<z<0.5$ \citep{muzzin2013evolution, behroozi2018universemachine, hahn2018iq, donnari2018star}, so the extrapolation to that regime needs to be tested with further data.

\subsection{The different demographics of galaxies}

\begin{figure*}
    \centering
    \includegraphics[width=500px]{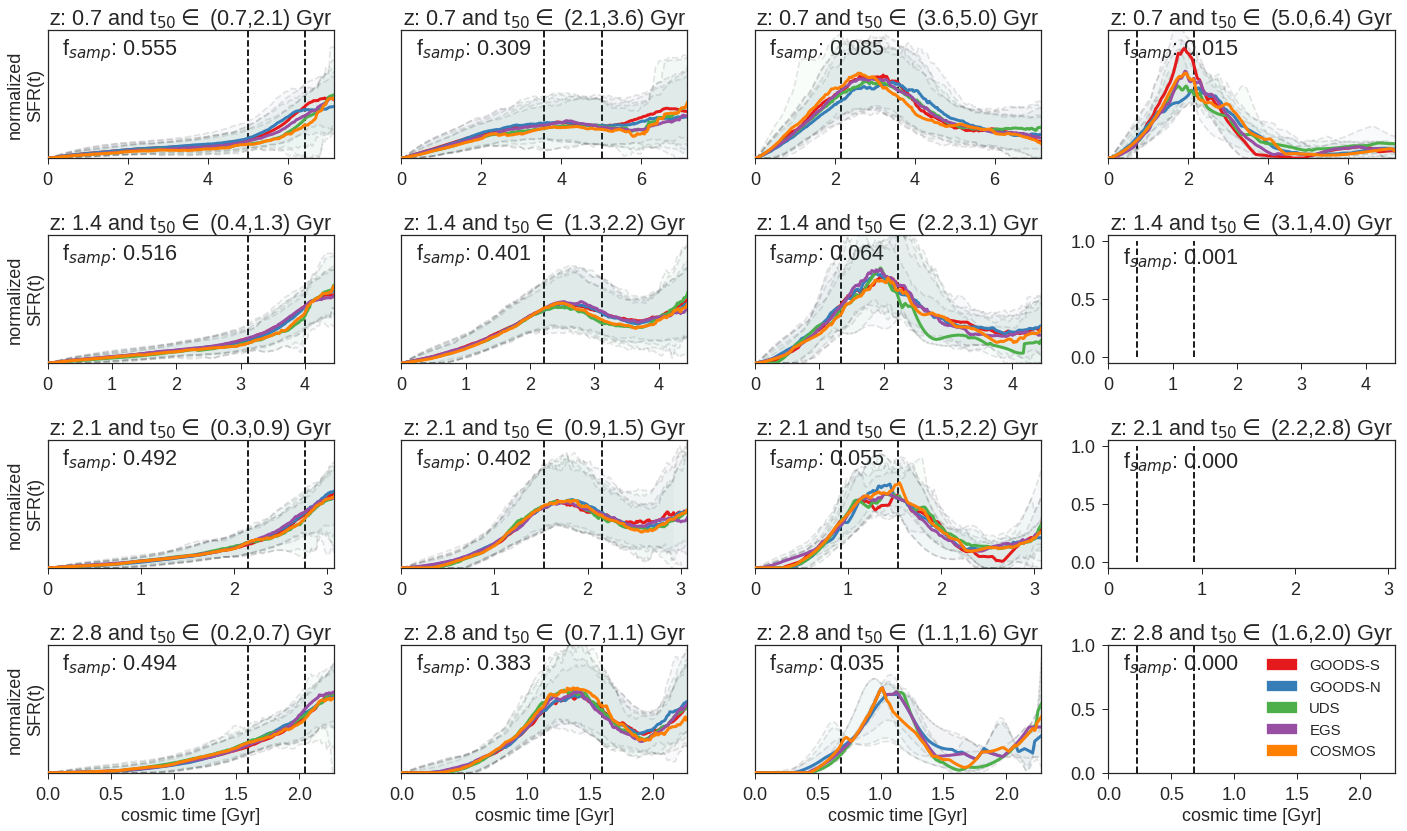}
    \caption{SFHs split into linearly increasing bins of $t_{50}$, the lookback time at which a galaxy assembled $50\%$ of its stellar mass, from $0.1 t_{univ} < t_{50} < 0.3t_{univ}$, $0.3t_{univ} < t_{50} < 0.5t_{univ}$, $0.5t_{univ} < t_{50} < 0.7t_{univ}$, and $0.7t_{univ} < t_{50} < 0.9t_{univ}$in four bins of redshift. The plotting scheme and colors are the same as Figure \ref{fig:sfh_all_mass_z}. In each bin, the SFHs are normalized to the same mass since we are most interested in the diversity of SFH shapes for the entire demographic. Vertical dashed black lines show the $t_{50}$ bounds for each panel. We see that the SFHs in a bin broadly tend to describe one of four demographics of galaxies: starbursting galaxies at high redshifts and late bloomers at $z\sim 0.7$  \cite{dressler2016demonstrating} can be found in the first column from the left, star forming galaxies contribute to the median SFH in columns 1-3, post-starburst or green-valley galaxies \citep{fang2012slow} in columns 2-3 and quiescent galaxies in columns 3-4. The $f_{samp}$ at the top left of each panel shows the fraction of galaxies at each redshift that fall into each demographic. In addition to the UVJ diagram and position on the SFR-M$_*$ plane, the SFHs of galaxies allow for additional diagnostics regarding its evolutionary phase.}
    \label{fig:sfh_all_t50lin_z}
\end{figure*}

\begin{figure*}
    \centering
    \includegraphics[width=500px]{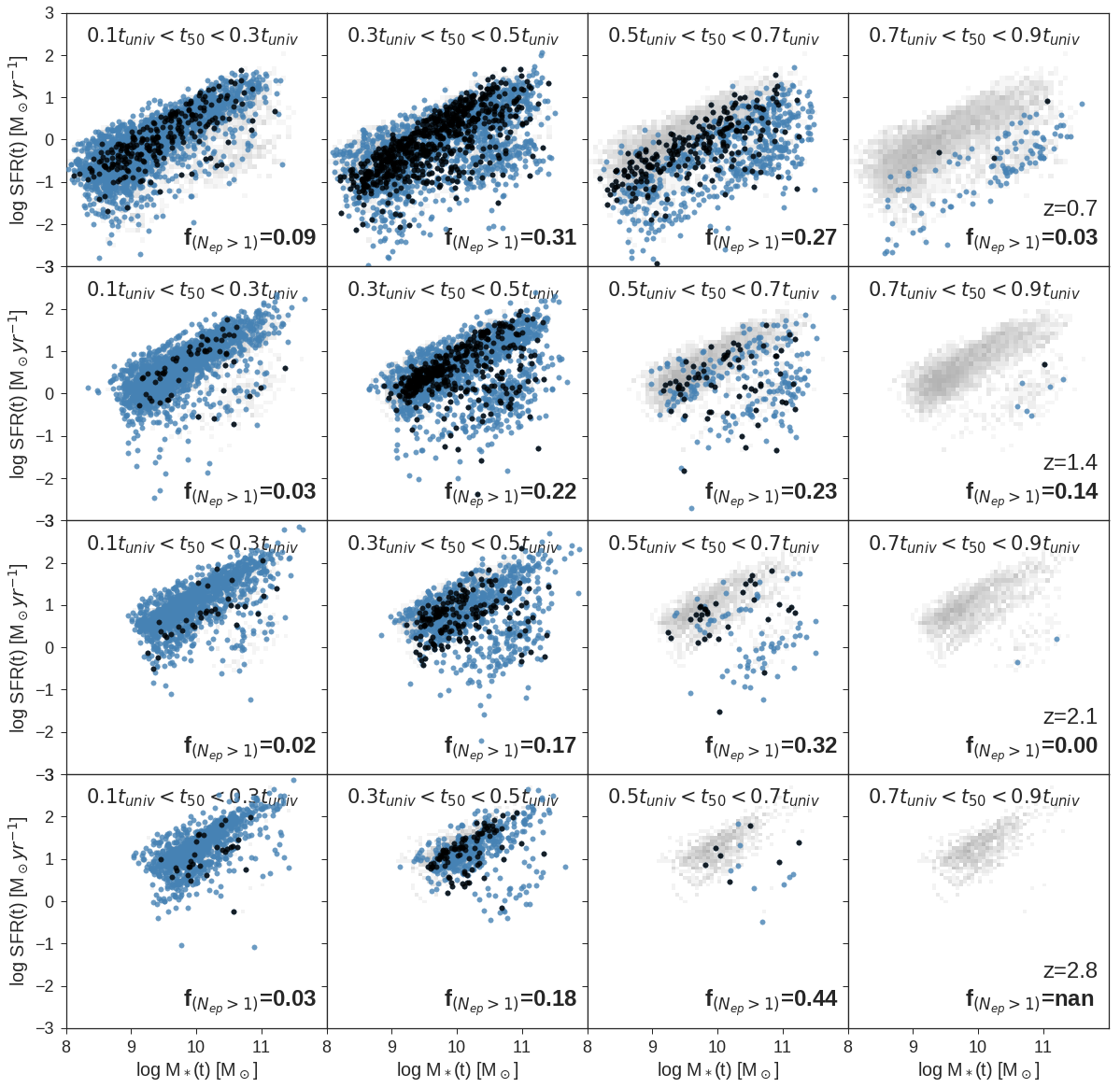}
    \caption{Positions on the SFR-M$_*$ plane for the galaxies shown in each panel of Figure. \ref{fig:sfh_all_t50lin_z} above. The underlying grey heatmap shows the full sample at a given redshift, and blue points show all galaxies satisfying  $0.1 t_{univ} < t_{50} < 0.3t_{univ}$, $0.3t_{univ} < t_{50} < 0.5t_{univ}$, $0.5t_{univ} < t_{50} < 0.7t_{univ}$, and $0.7t_{univ} < t_{50} < 0.9t_{univ}$ within a redshift bin. The black points are a subset of the blue points that are identified as having more than one major episode of star formation during their lifetimes ($N_{ep} > 1$), with this fraction of galaxies given in the bottom right corner. }
    \label{fig:sfr_mstar_all_t50lin_z}
\end{figure*}

In keeping with \citet{pacifici2016timing}'s finding that the widths of the SFHs of passive galaxies are roughly constant upon factoring out the age of the universe and our similar finding for the time interval between two peaks of star formation, we consider galaxy SFHs binned in $t_{50}$. We bin galaxies in four bins, from $0.1 t_{univ} < t_{50} < 0.3t_{univ}$, $0.3t_{univ} < t_{50} < 0.5t_{univ}$, $0.5t_{univ} < t_{50} < 0.7t_{univ}$, and $0.7t_{univ} < t_{50} < 0.9t_{univ}$ at different redshifts. We show the results in Figure \ref{fig:sfh_all_t50lin_z}. 
As in Figure \ref{fig:sfh_all_mass_z}, each panel lists the fraction of the sample in a particular bin. However in this case, the fractions are no longer tracing the mass function of galaxies. Instead, the four bins in time serve as proxies for galaxy SFHs in different stages of their lifetimes. This enables us to identify different populations of galaxies, including starbursting galaxies, late-bloomers \cite{dressler2016demonstrating}, star-forming galaxies, post-starburst or green-valley galaxies \citep{fang2012slow} and quiescent galaxies using different redshift-dependent $t_{50}$ cuts, either independently or in combination with other factors like $t_{25}, t_{75}$, size and morphology. To further interpret these SFHs,  Figure \ref{fig:sfr_mstar_all_t50lin_z} looks at the positions of the galaxies within each panel in Figure \ref{fig:sfh_all_t50lin_z} on the SFR-M$_*$ plane. We find that the left (right) columns in both figures select star forming (quiescent) galaxies that lie on (off) the star-forming sequence. At intermediate values of $t_{50}$, the populations are a combination of star forming  and quiescent galaxies, with the selection gradually shifting from star-forming to quiescent galaxies. The intermediate $t_{50}$ panels also show an excess of galaxies with multiple strong episodes of star formation ($N_{ep} > 1$). While this is an intuitive result, since galaxies that have assembled most of their mass recently or those that have long since shut off their star formation are not very likely to contain multiple episodes of star formation, it has important implications for analyses that assume that galaxies evolve along smooth SFH trajectories \citep{leitnerMSI} or are described by simple parametric forms \citep{dressler, ciesla2017sfr, lee2017intrinsic}.

\subsection{Correlation with morphology}

\begin{figure*}
    \centering
    \includegraphics[width=153px]{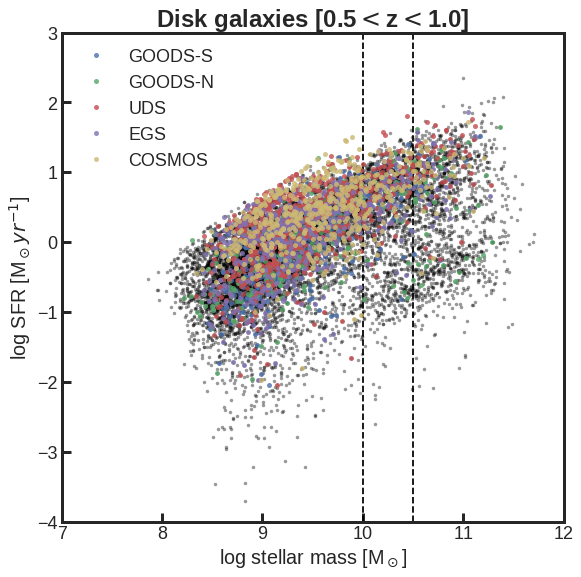}
    \includegraphics[width=153px]{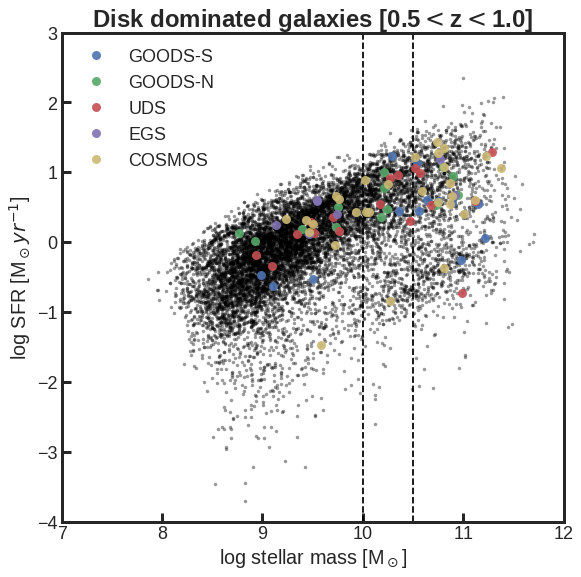}
    \includegraphics[width=153px]{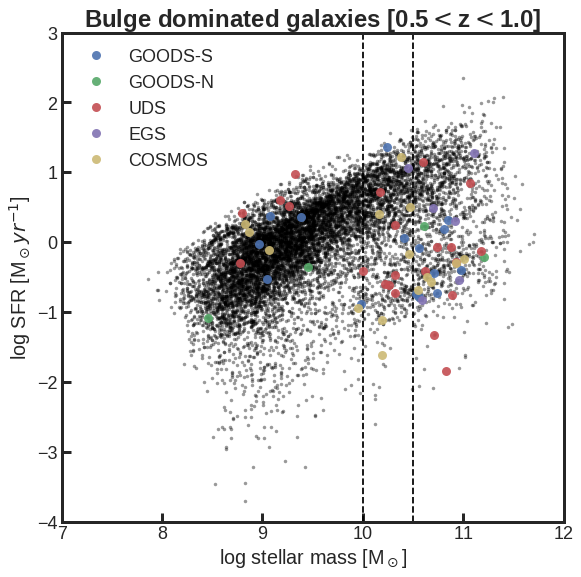}
    \includegraphics[width=150px]{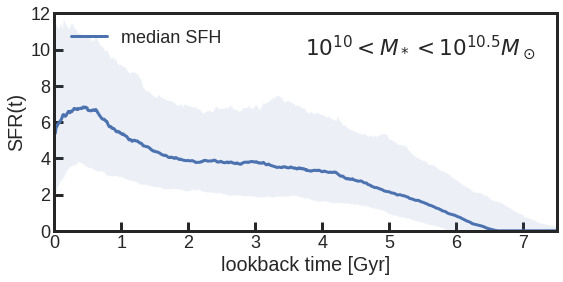}
    \includegraphics[width=150px]{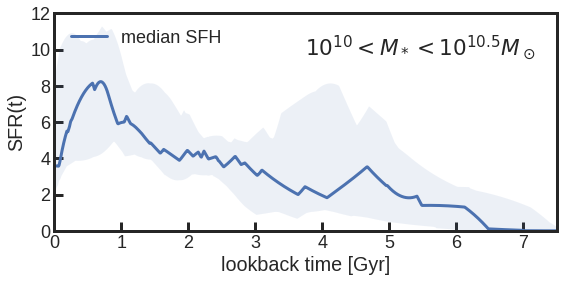}
    \includegraphics[width=150px]{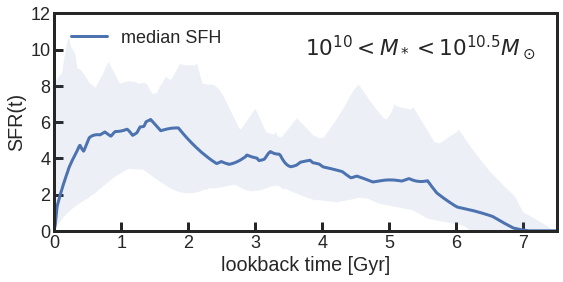}
    \includegraphics[width=153px]{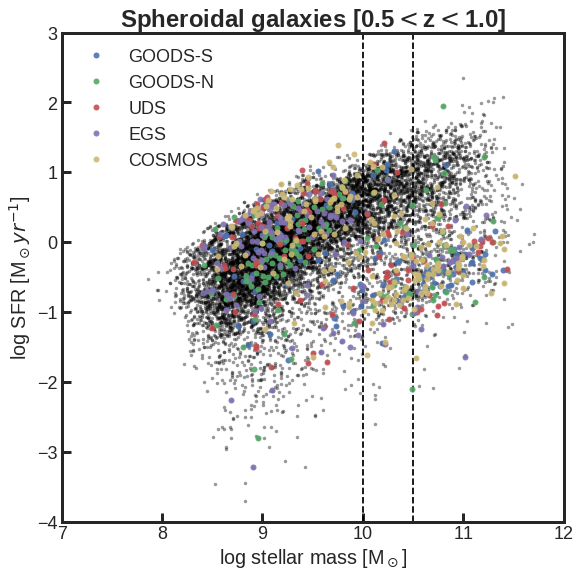}
    \includegraphics[width=153px]{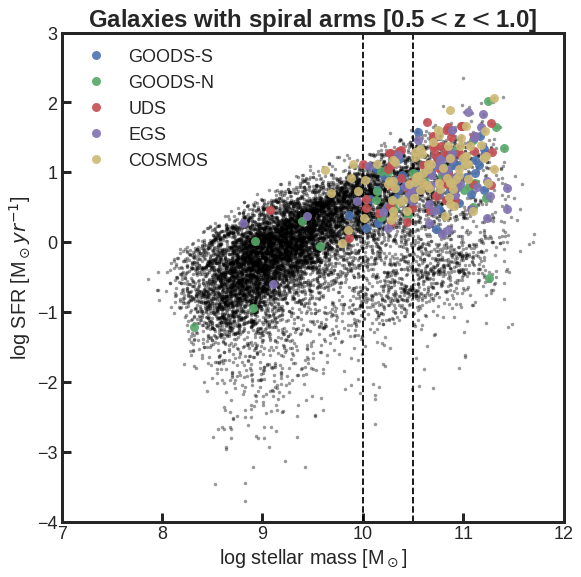}
    \includegraphics[width=153px]{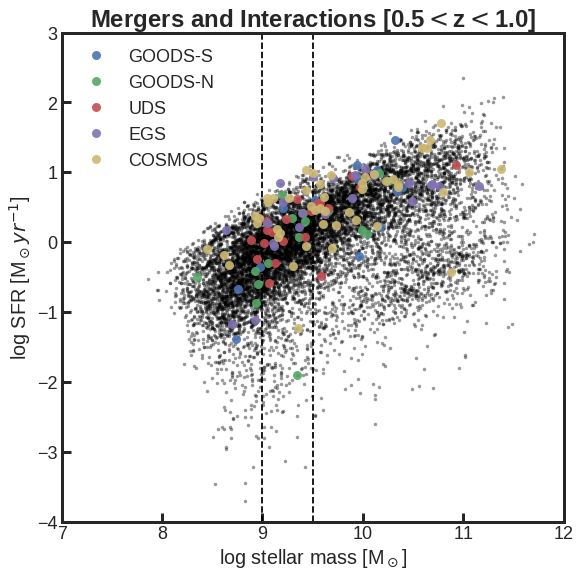}
    \includegraphics[width=150px]{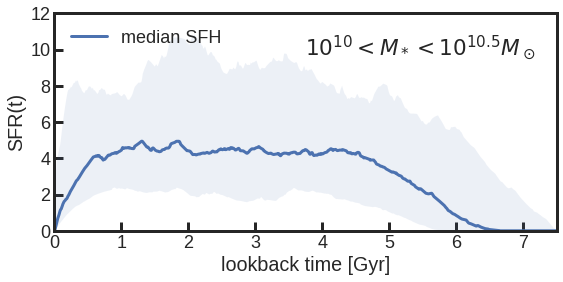}
    \includegraphics[width=150px]{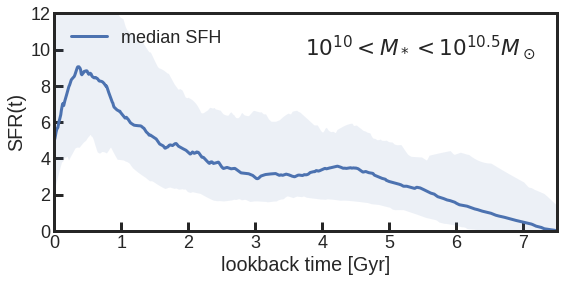}
    \includegraphics[width=150px]{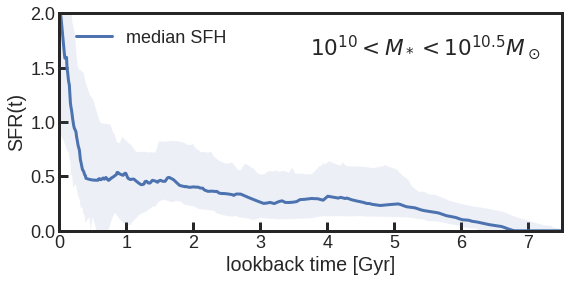}
    \caption{The star formation histories of galaxies at $0.5<z<1.0$ with different morphological features identified in \citet{kartaltepe2015candels}. For each class, the top panel shows the position of the galaxies under consideration (colored points) in the SFR-M$_*$ plane, with the full population shown as black dots. The bottom panel shows the median SFH (blue solid line) and diversity (16$^{th}$ to 84$^{th}$ percentile) shown as a shaded blue region, of all the galaxies with $M_* \in [10^{10},10^{10.5}] M_\odot$ that satisfy a particular morphological criterion, except for the last bin, where we have chosen a lower mass bin due to insufficient statistics. The second and third panels show galaxies that have both a disk and bulge component, which are then broken down into disk-dominated as opposed to bulge-dominated galaxies.}
    \label{fig:morph_sfhs}
\end{figure*}

The morphologies of galaxies are seen to strongly correlate with their stellar masses and redshifts \citep{conselice2014evolution}, as well as sSFR \citep{whitaker2015galaxy}. While this is a combined effect of the different processes that regulate star formation within galaxies, including mergers, gas accretion through inflows, stellar and AGN feedback \citep{hopkins2008cosmological, hopkins2014galaxies, angles2017black, weinberger2018supermassive, boselli2006environmental}, it is difficult to observationally disentangle the relative strengths of these effects. However, the different timescales that these processes act on enable us to discriminate between the relative effects of these processes if we can observationally constrain the timescales on which morphological transformation occurs across a population of galaxies, as was done for groups in eg. \citet{kovavc201010k}. The reconstructed SFHs of galaxies provide a direct probe of these timescales by correlating the morphologies of galaxies with their SFHs as compared to indirect measurements of timescales through the frequencies of different morphologies, which are subject to a variety of systematics and selection effects.

We use the CANDELS wide morphology catalogs by \citet{kartaltepe2015candels} to study the star formation histories of galaxies with different morphological features at $0.5<z<1.0$. We limit our redshift range to avoid the effects of small number statistics of classifications as we go to higher redshifts. The morphology catalogs contain visual classifications for over $50,000$ objects spanning $0<z<4$ with $f160w < 24.5$, which gives a large overlap with our sample. The catalogs contain flags for main morphology class (disk, spheroid, peculiar/irregular, point source/compact, and unclassifiable), a class for mergers and other interactions and structure flags for bars, tidal features, spiral arms and more. For each class and flag, the catalog reports the fraction of classifiers who were confident about the existence of that feature. 

We use this to analyse the SFHs of six classes of galaxies, described as follows:

\begin{itemize}
    \item \textbf{Disk:} ($f_{disk} > 0.9$) AND ($f_{sph}, f_{irreg} < 0.1$). This includes the set of all galaxies are classified as disky galaxies.
    \item \textbf{Disk dominated galaxies:} ($f_{disk~dom} > 0.9$). Disks with a central bulge where the disk dominates the structure.
    \item \textbf{Bulge dominated galaxies:} ($f_{bulge~dom} > 0.9$). Disks with a central bulge where the bulge dominates the structure.
    \textbf{Spheroid:} ($f_{sphk} > 0.9$) AND ($f_{disk}, f_{irreg} < 0.1$). This includes the set of all galaxies are classified as spheroidal galaxies.
    \item \textbf{Galaxies with spiral arms:} ($f_{arms} > 0.9$). 
    \item \textbf{Mergers and interactions:} galaxies that are either appear to have undergone a merger as evidenced by tidal features, structures such as loops or highly irregular outer isophotes ($f_{merger} > 0.9$) or are interacting with a companion galaxy within the segmentation map from SExtractor ($f_{int1} > 0.9$).
\end{itemize}

The results of this analysis are shown in Figure \ref{fig:morph_sfhs}. The Figure contains six sets of panels, one for each subsample of galaxies. The top panel shows where the galaxies lie on the SFR-M$_*$ correlation, and the bottom panel shows the median SFH for the subsample of these galaxies at $M_* \in [10^{10},10^{10.5}] M_\odot$. This is useful to test feedback driven models of quenching that posit a  correlation between bulge-total ratios and SFH shape \citep{zolotov2015compaction, tacchella2016confinement, belfiore2016sdss,abramson2018grism}. While the SFHs of our pure disk population at $M_*\sim 10^{10.25} M_\odot$seem to actively form stars throughout their lifetime, with the maximal peak in their SFHs maximum SFR close to the time of observation, the galaxies containing a disk and bulge component seem to show a downward trend in their median SFHs, with disk dominated galaxies peaking earlier on average than pure disks without a bulge, followed by a decline in SFR. This trend continues to bulge dominated galaxies and spheroids, showing an evolution in timescales that can be tested with simulations implementing different models for quiescence. For each population, we consider the lookback time at which the SFH of each galaxy peaked and use this distribution to quantify the timescale on which galaxy SFHs began their decline. For each population, we find the following (for galaxies at $0.5<z<1.0$ and $10^{10}<M_*<10^{10.5}M_\odot$, in lookback time):
\begin{itemize}
    \item Mergers: $t_{peak}: 0.00^{-0.00}_{+0.39}$ Gyr
    \item Galaxies with spiral arms: $t_{peak}: 0.60^{-0.54}_{+1.54}$ Gyr
    \item Disks: $t_{peak} = 0.81^{-0.80}_{+2.56}$ Gyr
    \item Disk-dominated galaxies: $t_{peak}: 0.70^{-0.38}_{+2.73}$ Gyr
    \item Bulge-dominated galaxies:$t_{peak}: 2.15^{-1.55}_{+3.07}$ Gyr
    \item Spheroids: $t_{peak}: 2.50^{-1.60}_{+2.25}$ Gyr
\end{itemize}

In the absence of a bulge component, we also see that galaxies with spiral arms inhabit the high-stellar mass, high SFR portion of the SFR-M$_*$ plane, continuing to actively form stars till they lose rotational support or start forming bulges. We also see that mergers and interactions show a noticeable increase in recent SFR, over timescales within the last $\sim 0.5$ Gyr of their SFHs. The timescales for these morphological transformations can be further constrained by determining the resolved SFHs of individual galaxies using IFU surveys like SDSS-IV MaNGA and CALIFA \cite{delgado2014star,belfiore2016sdss}.

\section{Discussion} \label{sec:discussion}

\subsection{Improvements from better datasets, models, and priors}
\label{sec:improvements_data}

Since nonparametric methods make no explicit assumption about the form of SFHs, they are only as good as the data being used for SED fitting. In this regard, there are three main avenues for data-driven improvement: better wavelength resolution, better wavelength coverage, and better S/N. Spectroscopy contains more information about stellar populations of different ages and metallicity as compared to broadband photometry, but often suffers from wavelength dependent flux calibration issues that need to be accounted for prior to fitting. Panchromatic SEDs allow us to test models of dust attenuation and re-emission to better constrain dust effects while estimating the SFHs of galaxies. 

The SFH reconstructions we obtain are also subject to several modeling uncertainties: Stellar Population Synthesis models can introduce systematics into the mapping between physical parameters and observed photometry \citep{conroy2010propagation, han2018comprehensive}. Differences in dust models can introduce systematics into the measurement of recent SFR, which would then propagate into differences in the SFH. Pacifici et al. (2019) in prep. compares the results from 14 different SED fitting codes applied to the same sample of CANDELS/GOODS-S galaxies at $z\sim 1$. This allows us to examine the effects of inter-code variability and model assumptions for dust, IMF, SFH, and SPS models including the effects of binary populations. Additionally, \citet{han2018comprehensive} tested multiple SPS models, SFH assumptions and dust models using a comprehensive bayesian formalism that allowed them to estimate the bayesian evidence in comparing different models. 

Finally, the choices of prior assumed during SED fitting are extremely influential in the estimates of physical parameters and their covariances. While we have tried to be agnostic about the priors in this work, it is important to note that an informative prior could be especially useful while fitting noisy, low S/N data with limited wavelength coverage. Predictive checks could be put in place to ensure that the priors do not introduce significant biases into the estimates, or cause artificially tight correlations due to regression to the mean. These informative priors could be developed by studying the distributions of physical quantities at a particular epoch from a small subset of high S/N observations, scaling relations and mass functions, as well as semi-analytic or empirical models that encode the physics that lead to these observables, explicitly quantifying the covariance between star formation, chemical attenuation and dust enrichment and destruction histories.

\subsection{SFHs as a probe to higher redshifts}

The SFHs of galaxies allow us to probe the behavior of mass functions and scaling relations out to higher redshifts than is currently possible. At low to intermediate redshifts, this can be used as a consistency check, to ensure that the reconstructed SFHs are not biased due to noise or prior assumptions. At high redshifts, this can be a powerful tool to increase observational statistics and push measurements out to higher redshifts that is directly accessible through observations. 

\citet{iyer2018sfr} propagated galaxies backwards in time along their SFHs in the form of trajectories in SFR-M$_*$ space to probe the high-redshift low stellar-mass regime of the SFR-M$_*$ correlation, finding that the projected correlation at intermediate redshifts matches the observed distribution well, and extending it by nearly two orders of magnitude out to $z \sim 6$ where observations are extremely faint. Pacifici et al. (2019), in prep. implements a validation test by reconstructing the stellar mass function using galaxies at lower redshifts and comparing them to the stellar mass function obtained through direct fits. \citet{leja2018quiescent} use star formation histories to probe the cosmic SFRD using galaxies at low redshifts, finding that the apparent mismatch between the mass functions and star formation rate functions is alleviated using nonparametric SFHs.

\subsection{Galaxy evolution studies enabled by SFH reconstruction}

The smooth, non-parametric star formation histories obtained with the improved Dense Basis method offer a window into the pasts of different galaxy populations. While interesting itself, this has the potential to be combined with a variety of ancillary data to probe a wide range of previously inaccessible quantities, some of which we briefly describe below:
\begin{itemize}
    \item Higher S/N observations or spectrophotometric data would help obtain better SFH constraints, allowing us to better constrain the number of major episodes of star formation and the timescales on which rejuvenation, starbursts, and quiescence occur at different epochs. While the current observations (this work, \citet{abramson2016return, pacifici2016timing}) show a flat trend with mass and a linear one with the age of the universe, it is an interesting problem to understand the physical mechanisms responsible for this trend and the dispersion of $\sim 0.25 t_{univ}$ Gyr using simulations.
    \item Spatially resolved SFHs computed using IFU data from surveys like SDSS-IV MaNGA and CALIFA allow us to better understand the correlation between the SFH and morphology and discriminate between inside-out vs outside-in scenarios for galaxy growth and quenching \citep{goddard2016sdss}, better examine the connection between the physical properties of individual regions within galaxies and their SFHs \citep{rowlands2018sdss} and test scaling relations at different regimes \cite{hsieh2017sdss}. Care needs to be exercised in interpreting these results since we only see where the stellar populations are today. Additional kinematic information would help alleviate this problem to a certain extent.
    \item Correlating SFHs with environment, size, kinematics, central density and morphology in addition to stellar mass, SFR and redshift could help build a unified picture of how galaxies evolve, with the SFHs providing a link between galaxy populations of different types and their earlier progenitors. Although this approach is similar to empirical models \citep{behroozi2018universemachine, moster2018emerge}, it has the advantage of much richer observational constraints from the individual SFHs of galaxies. A comparison of SFH distributions between simulations and observations would allow us to qualify additional factors that are not directly accessible.
    \item The Gaussian Process based parametrization can also be used as a general compression method for compressing and storing PDFs from all kinds of codes, similar to \citet{malz2018approximating}.
\end{itemize}

\subsection{Caveats in SED fitting and SFH reconstruction}

While we have performed an extensive range of validation tests (appendix \ref{sec:validation}) to ensure that all the quantities reported in this work are robust, there are some caveats to keep in mind while extending the SED fitting to different datasets or the analysis beyond what is performed here.

\begin{itemize}
    \item \textbf{SFHs are not mass accretion histories}: The SFH is a record of when the stars present in a galaxy at the time of observation were formed, as opposed to the mass accretion history, which is a record of when those stars entered the galaxy. These two quantities are the same for stars formed \textit{in-situ}, but are differ when the stars were brought in through mergers. This needs to be taken into account in certain kinds of analysis, for example by using a mass- and redshift-dependent merger fraction that to correct for mass functions calculated by propagating galaxies backwards in time along their SFHs. 
    \item \textbf{Lack of sensitivity to the shortest timescales}: The smooth SFHs reconstructed using SED fitting in this work can not capture starbursts that can happen on extremely short timescales of $\sim \mathcal{O}(10)$ Myr. While fitting galaxies from the semi-analytic model that contain such starbursts, we generally find that the overall stellar mass is well recovered, but the starburst is smeared out over larger timescales, depending on when it occured. This needs to be accounted for in the uncertainty budget for example, while calculating the scatter along the SFR-M$_*$ correlation using galaxies propagated backwards in time along their SFR-M$_*$ trajectories.
    \item \textbf{Non-uniform sensitivity to variations in SFH:} SED fitting is more sensitive to recent star formation than it is to star formation older than a few Gyr. As we go back in time, our SFH reconstruction transitions from being likelihood dominated to being prior dominated, and while we show that this does not cause biases in our SFH reconstruction in appendix \ref{sec:validation}, it does mean that we are less sensitive to sharp variations in SFH at large lookback times compared to closer to the time of observation.
    \item \textbf{Correlation with chemical enrichment histories:} While we have considered the problem of estimating the star formation histories of galaxies in this work, in practice they are highly correlated with the chemical enrichment histories of galaxies. While metallicity is poorly constrained in our current observations, while working with higher S/N data or spectra the analysis should include a joint model for SFR(t) and Z(t), which can be achieved through joint priors on the metallicity given by Z($M_*, \{t_x\}$) informed using simulations. 
\end{itemize}

\section{Conclusions}
\label{sec:conclusions}

Studying the star formation histories (SFHs) of galaxies lets us better understand the timescales on which different physical processes shape galaxy growth. High S/N multiwavelength observations from current and upcoming galaxy surveys make it possible to reconstruct the SFHs for large ensembles of galaxies with suitably sophisticated analysis techniques.

We update the Dense Basis Spectral Energy Distribution (SED)-fitting method \citep{iyer2017reconstruction} using a flexible SFH parametrization described by the tuple (M$_*$, SFR, $\{ t_x \}$) where M$_*$ is the stellar mass, SFR is the star formation rate averaged over the past 100 Myr, and the set $\{ t_x \}$
contains the lookback times at which a galaxy formed N equally spaced quantiles of its stellar mass. These parameters represent a set of integral constraints and SFHs corresponding to a particular tuple are constructed using Gaussian Process regression in fractional mass-cosmic time space, which creates smooth curves that satisfies these constraints and is completely independent of the choice of a functional form. We reconstruct the SFHs of galaxies with uncertainties using a brute-force bayesian approach with a large pre-grid of model SEDs. To make the method fully nonparametric, we determine N on an SED to SED basis using a Bayesian Information Criterion (BIC) based selection. Using the reconstructed SFHs and a peak finding algorithm, we determine the number of major episodes of star formation in a galaxy's past. 

The method provides the following advantages:    

\begin{itemize}
    \item The method encodes the maximal amount of SFH information in a minimal number of parameters.
    \item Being independent of the choice of a functional form, it does not suffer from the traditional biases associated with simple parametric assumptions for the SFH shape. 
    \item The method also circumvents the pitfalls associated with the traditional nonparametric approach of describing SFHs as fixed bins in lookback time with constant SFR within a bin such as artifacts due to bin edges and reduces uneven S/N distribution across different parameters.
    \item The parameters used to describe SFHs are physically interpretable, and allow eazy comparison between different datasets from observations and simulations. 
    \item Informative priors can be constructed by studying these parameters in cosmological simulations and semi-analytic models, which can be used while fitting low S/N data or SEDs with partial wavelength coverage.
    \item The method is computationally fast, able to fit $\sim 34$ galaxies /minute/core on a 2.9 GHz Intel processor, and capable of being adapted to most data compression problems. 
\end{itemize}

We apply the method to a sample of 48,791 galaxies across the five CANDELS fields with HST/WFC3 $F160W<25$ and $0.5<z<3.0$. 

We use the reconstructed SFHs to study galaxy evolution across stellar mass and redshift, and quantify the fraction of galaxies at each epoch that have multiple strong episodes of star formation. For the galaxies that show multiple strong episodes of star formation, we find that timescale separating two peaks in the SFH is roughly constant with mass, and increases linearly with the age of the universe as $t_{peak-to-peak} \sim 0.42_{-0.10}^{+0.15} t_{univ}$ Gyr. We also find that classifying galaxies by $t_{50}$ is a robust way of selecting for star forming galaxies at a given epoch.

Using the \citet{kartaltepe2015candels} morphology catalog, we can examine the SFHs for subsets of galaxies with particular morphological features, finding the expected correlation between the SFHs of galaxies and morphological features. In addition, we quantify the timescale on which the SFH declines as a function of morphology, finding that this increases from $\sim 0.60^{-0.54}_{+1.54}$ Gyr for galaxies with spiral arms to $2.50^{-1.60}_{+2.25}$
Gyr for spheroids.    
    
The SFH formalism presented here is broad in scope and can be incorporated into any SED fitting code, can be used to compress and store SFHs in simulations, and can be used as a common parametrization to compare SFHs across different observations and simulations.

\section*{Acknowledgements}

The authors would like to thank Louis Abramson and Steven Finkelstein for their insightful comments and suggestions. KI would like to thank Boris Leistedt for a wonderful talk introducing Gaussian Processes.
The authors acknowledge Dritan Kodra, Jeff Newman, Steve Finkelstein, Adriano Fontana, Janine Pforr, Mara Salvato, Tommy Wiklind, and Stijn Wuyts for generating the photo-z PDFs for the compilation of $z_{\rm best}$ in the v2 CANDELS photo-z catalog, Guillermo Barro for compiling the GOODS-North photometric catalog, and Viraj Pandya for compiling the Santa Cruz semi-analytic model SFHs used in this work. KI \& EG gratefully acknowledge support from Rutgers University. 
This work used resources from the Rutgers Discovery Informatics Institute, supported by Rutgers and the State of New Jersey. 
The Flatiron Institute is supported by the Simons Foundation.
Support for Program number HST-AR-14564.001-A and GO-12060 was provided by NASA through a grant from the Space Telescope Science Institute, which is operated by the Association of Universities for Research in Astronomy, Incorporated, under NASA contract NAS5-26555.

\bibliography{db_refs.bib} 


\appendix

\section{Validation tests}
\label{app:validation} 

\subsection{SFH robustness}
\label{app:SFH_validation}

\begin{figure}[h]
    \centering
    \includegraphics[width=250px]{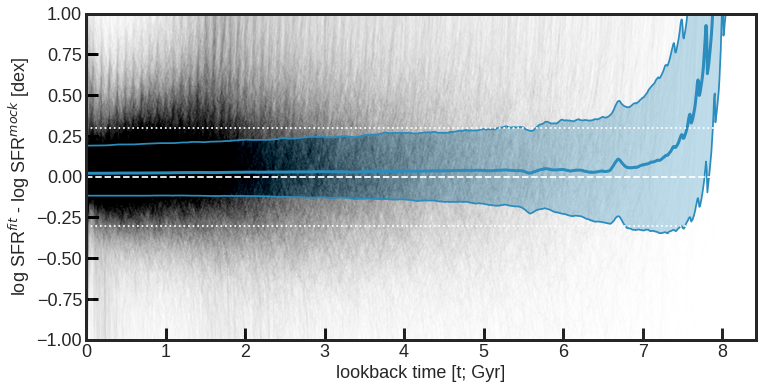}
    \includegraphics[width=250px]{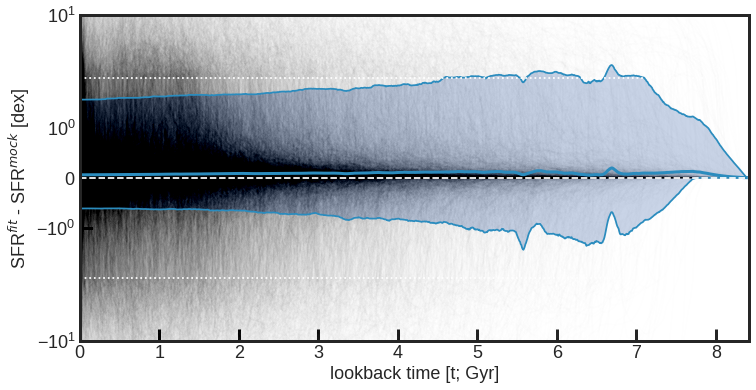}
    \caption{Comparing the ensemble of true SFHs from cosmological simulations to SFHs reconstructed from SEDs using the Gaussian Process Dense Basis method. Thin black lines show the difference in log SFR for individual galaxies, with the pointwise median in time shown as a solid blue line and the shaded blue region denoting the 16th to 84th percentile. The left panel shows the difference in terms of $\Delta \log SFR$, while the right panel shows the difference in terms of $\Delta SFR$ at each point in lookback time.}
    \label{fig:sfh_validation}
\end{figure}

To quantify possible biases in our SFH reconstruction, we plot the difference between the true and recovered SFH for our sample of validation galaxies in Figure \ref{fig:sfh_validation}. The two panels show the linear and logarithmic differences between the true and the recovered SFHs, in both cases this difference is smaller than 0.3 dex for most of cosmic time. While the reconstructed SFHs can not recover every short episode of star formation, as seen in the errors for individual galaxies (thin black lines), the overall SFH for the ensemble of galaxies is unbiased out to nearly $\sim 8 Gyr$. In the $\Delta \log SFR$ error plot, the bias blows up as we approach the big bang. This is because the SFHs in the mocks can abruptly fall to 0 close to the big bang while the Gaussian Process SFHs smoothly decline to 0 at $t=0$, leading to a one sided error. However, the arithmetic $\Delta SFR$ plot shows that this is in fact a very small difference, exaggerated by the fact that log $SFR_{true} \to -\infty$ as $SFR_{true}\to 0$ close to the big bang.

\subsection{Parameter robustness}
\label{app:param_validation}

\begin{figure}[h]
    \centering
    \includegraphics[width=490px]{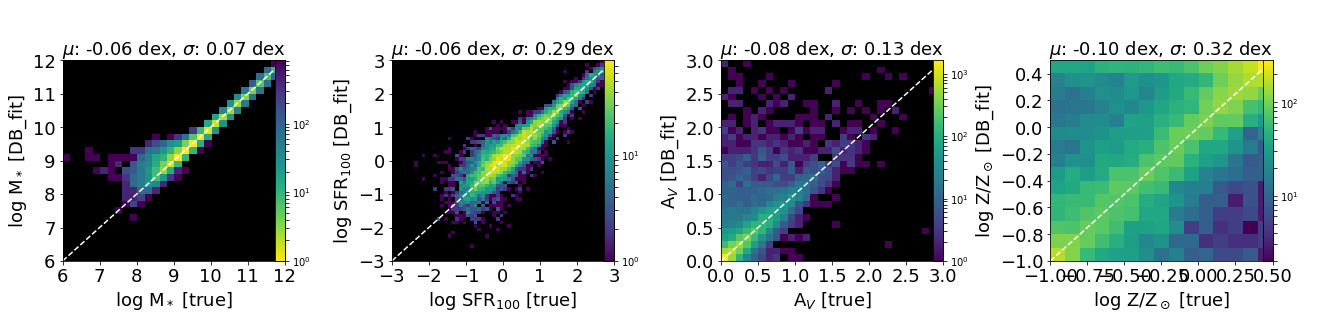}
    \caption{Results of estimating stellar masses, star formation rates, dust attenuation and metallicities for the ensemble of mock galaxies described in Appendix \ref{sec:validation}. Each parameter is shown as a log-scale heatmap, with the adjacent colorbar detailing the number of galaxies in a given bin. Bins with no galaxies are shown in black. The dashed white line shows the 1:1 relation, and each plot title contains the overall bias and scatter around the mean for the sample.}
    \label{fig:validation_ensemble}
\end{figure}

In addition to the star formation histories, Figure \ref{fig:validation_ensemble} shows the results of estimating traditional SED fit parameters - stellar mass, star formation rate averaged over the last 100 Myr, dust attenuation and stellar metallicity for the validation catalog. The stellar masses are constrained  better than the traditional scatter around the mean of $\sim 0.14$ dex \citep{mobasher2015critical}. Star formation rates have a larger scatter (0.29 dex) due to degeneracies with dust attenuation (0.13 dex) and metallicity (0.32 dex). At low stellar masses, we are prone to overestimating the overall mass due to our choices of SFH prior dominating the fit for low S/N SEDs. Restricting the fits to galaxies with $H < 25$ is found to largely exclude these low S/N objects, leading to more robust estimates of these parameters.

\subsection{SFH uncertainties}
\label{app:uncert_validation}

\begin{figure}[h]
    \centering
    \includegraphics[width=150px]{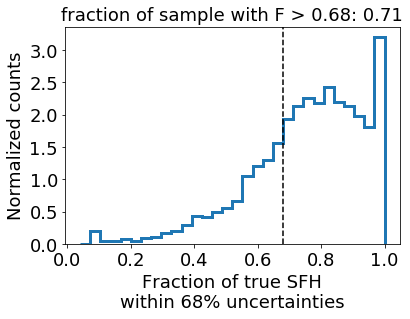}
    \includegraphics[width=150px]{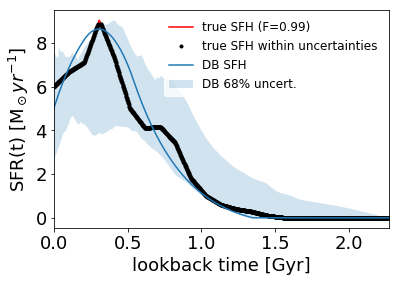}
    \includegraphics[width=150px]{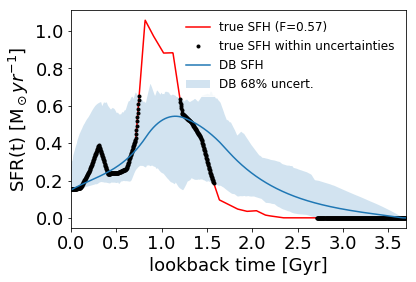}
    \caption{Validation performed to check the robustness of SFH uncertainties estimated from using the posterior from SED fitting. For our sample of 10,000 mock galaxies, we compute the fraction of the true SFH that lies within the 68\% uncertainties, shown as a histogram in the left panel, with a black dashed line indicates 0.68. We find that this condition is satisfied for 71\% of the sample, indicating that our uncertainties are robust. The middle and right panels show two examples of this computation for two galaxies in our sample, one where the majority of the true SFH lies within the uncertainties (middle) and the other where a sharp peak lies outside the 68\% uncertainties (right).}
    \label{fig:uncert_validation}
\end{figure}

As described in Sec. \ref{sec:method}, SFH uncertainties are computed at each point in lookback time as the $16^{th}$ to $84^{th}$ percentiles of SFHs constructed from 100 draws from the SFH posterior ($M_*, SFR, \{t_x\}$) using the Gaussian Process routine. To check whether these represent true 68$\%$ uncertainties, we consider the uncertainties estimated for each of the 10,000 galaxies in the mock catalog that we fit and compute the fraction of the true SFH that lies within the uncertainties for each galaxy. The distribution of this fraction is given in the left panel of Figure \ref{fig:uncert_validation}, which shows that the truth lies within the uncertainties $\sim 71\%$ of the time. This indicates that our uncertainties are robust, similar to \citet{iyer2017reconstruction}. 
Figure \ref{fig:uncert_validation} shows two examples, one  where the uncertainties are representative of the truth, and another where both the reconstruction and the uncertainties miss a relatively short ($< 0.5 Gyr$) episode of star formation that follows a relatively quiescent period, due to the smoother reconstructed SFH producing a comparable SED given the photometric noise. 

\subsection{Number of episode estimation}
\label{app:nep_validation}

\begin{figure}[h]
    \centering
    \includegraphics[width=250px]{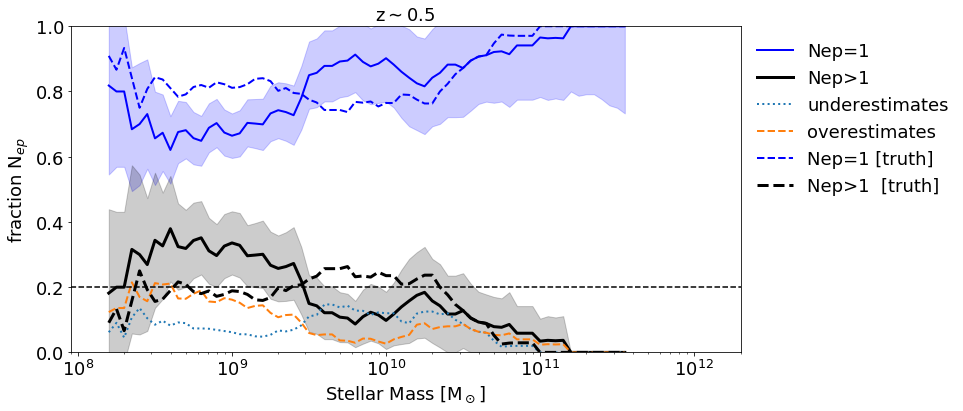}
    \includegraphics[width=250px]{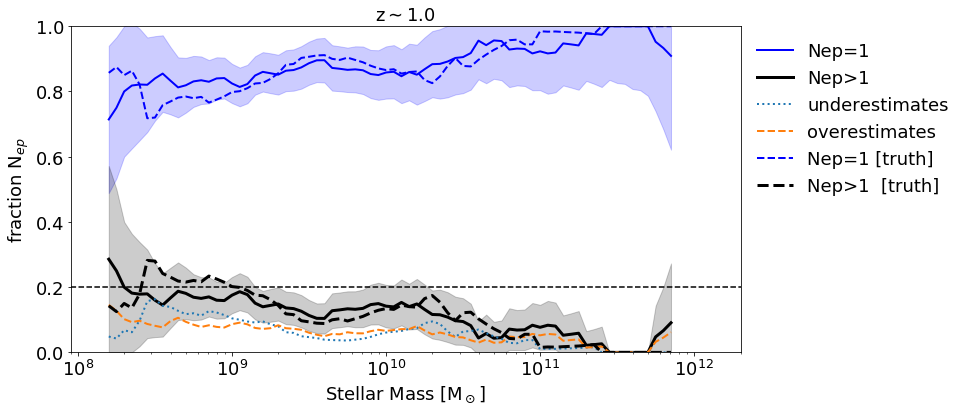}
    \includegraphics[width=250px]{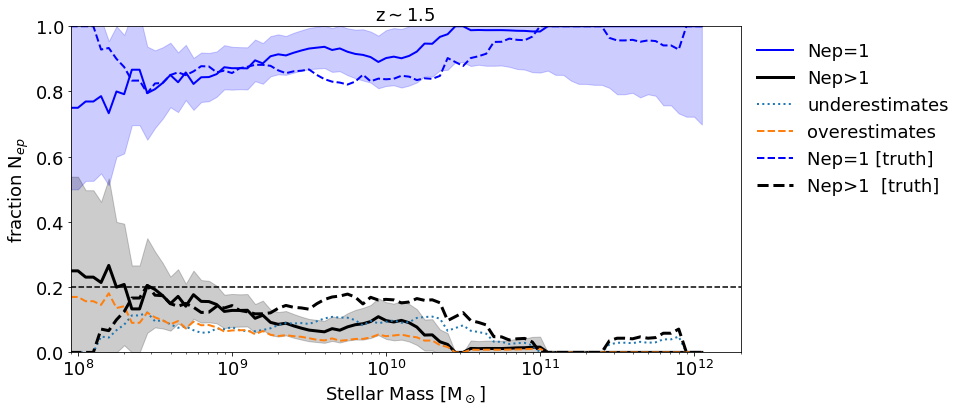}
    \includegraphics[width=250px]{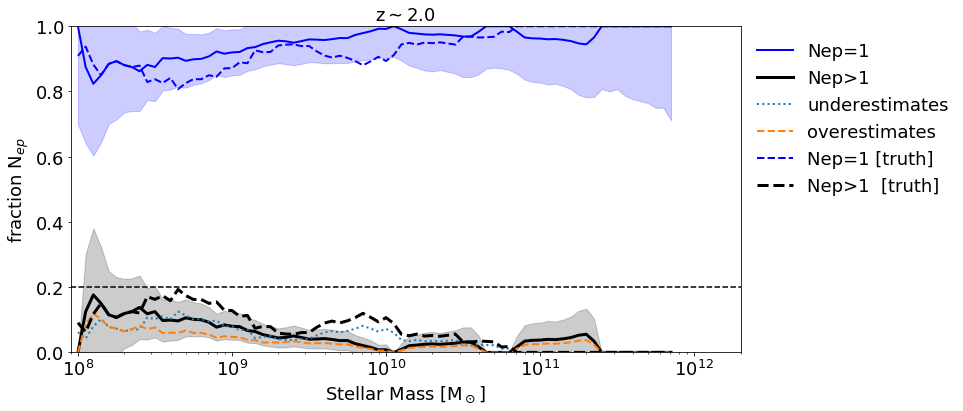}
    \includegraphics[width=250px]{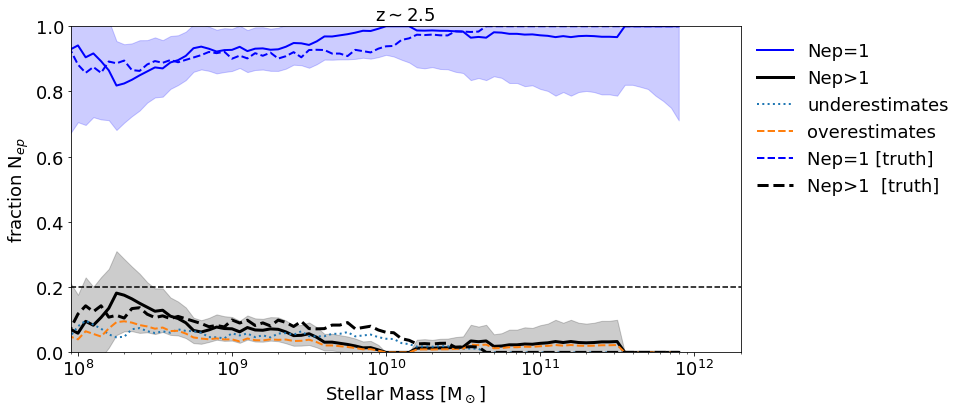}
    \includegraphics[width=250px]{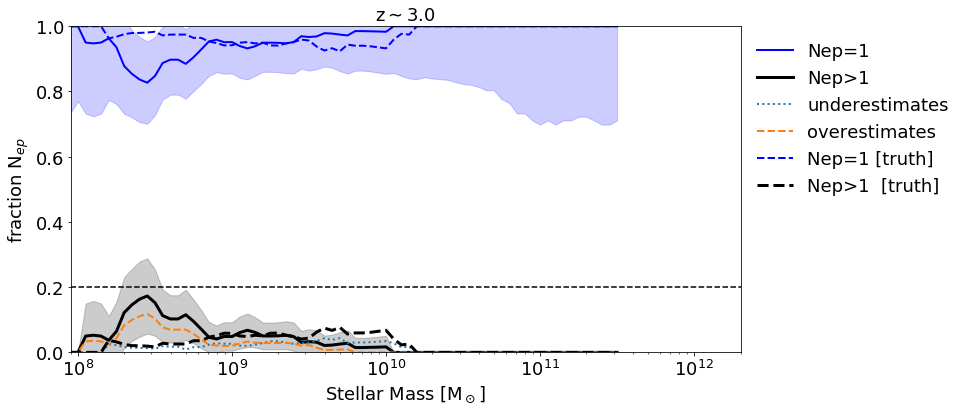}
    \caption{Validation tests recovering the number of major episodes of star formation in a galaxy's past by fitting the ensemble of mock noisy SEDs described in App. \ref{sec:validation}. The plots show the results of our analysis for the same redshift bins and mass range used for the main CANDELS sample. The solid lines show a sliding median within $\pm 0.25$ dex in stellar mass for each quantity, the shaded regions show the uncertainty for the estimates assuming Poisson noise}
    \label{fig:validation_nep}
\end{figure}

For each galaxy in the mock sample, we compute the number of major episodes of star formation ($N_{ep}$) using the peak-finding algorithm described in Sec. \ref{sec:method}. For the true SFHs, we compute the number of episodes requiring a dip of $log SFR_{peak} - log SFR_{min,local} \geq 0.3$ dex, to model the scenario of a galaxy on the star-forming sequence dropping below the sequence before rejoining it. The distribution of galaxies with multiple strong episodes of star formation for the mock sample is then given by the thick dashed black lines in Figure \ref{fig:validation_nep}. We then estimate the number of episodes for the reconstructed SFHs with the same criterion, and find a flatter trend than observed in the mocks. This is a combination of two effects: (1) The S/N contributed to the overall SED by a stellar population of a certain age decreases roughly logarithmically with lookback time \citep{ocvirk2006steckmap} leading to poorer constraints on intermediate and older stellar populations. (2) There is a mass-dependent correlation towards older stellar populations as galaxies grow more massive. As a result of this, galaxies that formed most of their mass, followed by an extended period of quiescence can be erroneously classified as having multiple episodes since when the SFR at the time of observation is low, it is easier to have a fluctuation of $\geq 0.3$ dex with a small variation in SFR.
We find that introducing a stellar mass dependent threshold to determine the number of peaks given by equation \ref{eqn:sfr_threshold} accounts for these mass-dependent effects on the distribution, allowing us to better reproduce the true trends in $N_{ep}$ with stellar mass and redshift, as seen in Figure. \ref{fig:validation_nep}. The thin lines also show the number of overestimates and underestimates, such that $N_{ep, true} = N_{ep, rec}$ - underestimates + overestimates. Since we are using the best-fit SFHs for this computation, we find that our results are quite sensitive to noise for individual galaxies. However, the distributions are estimated robustly, across a range of stellar masses and redshifts. Additionally, the differing behaviour of the observational sample from our mock catalog is indicative of the fact that our mass-dependent thereshold does not impose an artificial trend on the distribution. Reconciling the differences between the two distributions is an interesting topic for further study.




\end{document}